\documentclass{ws-ijmpa}
\usepackage[super,compress]{cite}
\usepackage{graphicx}
\usepackage{graphicx}
\usepackage{dcolumn}
\usepackage{bm}
\usepackage{epstopdf}
\newcommand{\hs}{\hspace*{0.5cm}}

\newcommand{\be}{\begin{equation}}
\newcommand{\ee}{\end{equation}}
\newcommand{\bea}{\begin{eqnarray}}
\newcommand{\eea}{\end{eqnarray}}
\newcommand{\ben}{\begin{enumerate}}
\newcommand{\een}{\end{enumerate}}
\newcommand{\bde}{\begin{widetext}}
\newcommand{\ede}{\end{widetext}}
\newcommand{\nn}{\nonumber}
\newcommand{\crn}{\nonumber \\}

\newcommand{\al}{\alpha}
\newcommand{\la}{\lambda}

\newcommand{\om}{\omega}

\newcommand{\fr}{\frac}
\newcommand{\bc}{\begin{center}}
\newcommand{\ec}{\end{center}}

\newcommand{\de}{\delta}
\newcommand{\De}{\Delta}

\newcommand{\varep}{\varepsilon}

\newcommand{\La}{\Lambda}
\newcommand{\si}{\sigma}

\newcommand{\bit}{\begin{itemize}}
\newcommand{\eit}{\end{itemize}}
\begin{document}
%
\catchline{}{}{}{}{}

\title{Neutrino mixing with non-zero $\theta_{13}$ and CP violation in the 3-3-1
model based on $A_4$ flavor symmetry}

\author{VO VAN VIEN }

\address{Department of Physics, Tay Nguyen University, 567 Le
Duan, Buon Ma Thuot, Vietnam\\
wvienk16@gmail.com}

\author{HOANG  NGOC  LONG}

\address{Institute of Physics,
VAST, 10 Dao Tan, Ba Dinh, Hanoi, Vietnam\\
hnlong@iop.vast.ac.vn}

\maketitle
\begin{history}
\received{Day Month Year}
\revised{Day Month Year}
\end{history}

\begin{abstract}
We propose a 3-3-1 model with neutral fermions based on $A_4$ flavor symmetry
responsible for fermion masses and mixings with non-zero $\theta_{13}$. To get realistic neutrino mixing,
we just add a new $SU(3)_L$ triplet being  in $\underline{3}$ under $A_4$. The neutrinos get
small masses from two $SU(3)_L$ antisextets and one $SU(3)_L$ triplet. The model can
 fit the present data on
neutrino masses and mixing as well as the effective mass governing neutrinoless double beta decay.
Our results show that the neutrino masses are naturally small and a little deviation from the
tri-bimaximal neutrino mixing form can be realized. The Dirac CP violation phase $\delta$ is predicted to either
  $5.41^{\circ}$ or $354.59^{\circ}$ with $\theta_{23} \neq  \fr \pi 4$.

\end{abstract}


\section{Introduction}
\label{intro}
Despite the great success of the Standard Model (SM)
of the elementary particle physics, the origin of flavor
structure, masses and mixings between generations of matter
particles are still open questions. The neutrino mass and mixing is one of
the most important evidence of beyond Standard Model physics. Many
experiments show that neutrinos have tiny masses and their mixing
is sill mysterious\cite{altar1, altar2} . The tri-bimaximal form for
explaining the lepton mixing scheme was first proposed by
Harrison-Perkins-Scott (HPS), which apart from the phase
redefinitions, is given by \cite{hps1, hps2, hps3, hps4}
\be
U_{\mathrm{HPS}}=\left(
\begin{array}{ccc}
\frac{2}{\sqrt{6}}       &\frac{1}{\sqrt{3}}  &0\\
-\frac{1}{\sqrt{6}}      &\frac{1}{\sqrt{3}}  &\frac{1}{\sqrt{2}}\\
-\frac{1}{\sqrt{6}}      &\frac{1}{\sqrt{3}}  &-\frac{1}{\sqrt{2}}
\end{array}\right),\label{Uhps}
\ee
can be considered  as a good approximation for the recent
neutrino experimental data. In fact, the absolute
values of the entries of the lepton mixing matrix $U_{PMNS}$ approximately are given by \cite{Uij,Uij1,Uij2,Uij3}
\be
\left|U_{\mathrm{PMNS}}\right|=\left(
\begin{array}{ccc}
0.795-0.846     &0.513-0.585  &0.126-0.178\\
0.205-0.543     &0.416-0.730  &0.579-0.808\\
0.215-0.548     &0.409-0.725  &0.567-0.800
\end{array}\right),\label{Uij}
\ee
The data in Refs. \citen{PDG2012,PDG1,PDG2,PDG3,PDG4} imply \bea
&&\sin^2(2\theta_{12})=0.857\pm 0.024,\,\,\sin^2(2\theta_{23})>
0.95,\,\,\,\sin^2(2\theta_{13})=0.098\pm 0.013,\crn && \Delta
m^2_{21}=(7.50\pm0.20)\times 10^{-5} \mathrm{eV}^2,\,\,\, \De
m^2_{32}=(2.32^{+0.12}_{-0.08})\times
10^{-3}\mathrm{eV}^2.\label{PDG2012}\eea Whereas, the best fit
values of neutrino mass squared differences and the leptonic
mixing angles in Refs. \citen{Nuexp1,Nuexp2} have been given to be slightly
modified from (\ref{PDG2012}), as shown in Tables \ref{NormalH}
and \ref{InvertedH}. 
\begin{table}[h]
\tbl{The experimental values of neutrino mass
squared splittings and leptonic mixing parameters, taken from
Refs. \citen{Nuexp1,Nuexp2} for normal hierarchy.}
{\begin{tabular}{@{}cccc@{}} \toprule
Parameter & Best fit & $1\sigma $ range& $2\sigma $ range \\
\noalign{\smallskip}\hline\noalign{\smallskip}
$\De m_{21}^{2}$($10^{-5}$eV$^2$) & $7.62$ & $7.43-7.81$ & $7.27-8.01$ \\
$\De m_{31}^{2}$($10^{-3}$eV$^2$)& $2.55$ & $2.64-2.61$ & $2.38-2.68$ \\
$\sin ^{2}\theta _{12}$ &  $0.320$ & $0.303-0.336$ & $0.29-0.35$   \\
 $\sin ^{2}\theta _{23}$&  $0.613$ & $0.573-0.635$ & $0.38-0.66$ \\
 $\sin ^{2}\theta_{13}$& $0.0246$ & $0.0218-0.0275$ & $0.019-0.03$  \\ \botrule
\end{tabular} \label{NormalH}}
\end{table}
\begin{table}[h]
\tbl{The experimental values of neutrino mass
squared splittings and leptonic mixing parameters, taken from
Refs. \citen{Nuexp1,Nuexp2} for inverted hierarchy.}
{\begin{tabular}{@{}cccc@{}} \toprule
Parameter & Best fit & $1\sigma $ range& $2\sigma $ range \\
\noalign{\smallskip}\hline\noalign{\smallskip}
$\De m_{21}^{2}$($10^{-5}$eV$^2$) & $7.62$ & $7.43-7.81$ & $7.27-8.01$ \\
$\De m_{13}^{2}$($10^{-3}$eV$^2$)& $2.43$ & $2.37-2.50$ & $2.29-2.58$ \\
$\sin ^{2}\theta _{12}$ &  $0.32$ & $0.303-0.336$ & $0.29-0.35$  \\
 $\sin ^{2}\theta _{23}$&  $0.60$ & $0.569-0.626$ & $0.39-0.65$ \\
 $\sin ^{2}\theta_{13}$& $0.025$ & $0.0223-0.0276$ & $0.02-0.03$\\ \botrule
\end{tabular} \label{InvertedH}}
\end{table}
These large neutrino mixing angles are completely different from the quark
 mixing ones defined by the Cabibbo- Kobayashi-Maskawa (CKM) matrix \cite{CKM, CKM1} .
 This has stimulated works on flavor symmetries and non-Abelian discrete
 symmetries, which are considered to be the most attractive candidate to
formulate dynamical principles that can lead to the flavor mixing
patterns for quarks and leptons. There are many recent models based
on the non-Abelian discrete symmetries, see for example \cite{dlshA4,dlsvS4,dlnvS3,vlD4,vlS4,vlS3,vlT7,vD4} and the references there in.

An alternative  extension of  the SM is the 3-3-1 models, in which
  the SM gauge group $SU(2)_L\otimes U(1)_Y$ is extended to
  $SU(3)_L\otimes U(1)_X$, whose phenomenology has been studied in great detail from various
particle physics standpoints \cite{331m1, 331m2, 331m3, 331m4, 331m5,331r1, 331r2, 331r3, 331r4, 331r5,e3311,e3312,e3313,e3314} .
 The anomaly
cancelation and the QCD asymptotic freedom  in the models require
that the number of  families is equal to the  number of quark
colors, and one family of quarks has to transform under
$\mathrm{SU}(3)_L$ differently from the two others. In our
previous works \cite{dlshA4,dlsvS4,dlnvS3,vlD4,vlS4,vlS3,vlT7,vD4} , the
discrete symmetries have been explored to the 3-3-1 models. The
simplest explanation is probably due to a $S_3$ flavor symmetry
which is the smallest non-Abelian discrete group,  has been
explored in our previous work \cite{dlnvS3} .
 In Refs. \citen{dlshA4,dlsvS4} we have studied the 3-3-1 model with neutral leptons based
    on $A_4$ and $S_4$ groups, in which the exact tri-bimaximal form is obtained,
     where $\theta_{13}= 0$.
As we know, the recent considerations have implied $\theta_{13}\neq 0$,
but relatively small as given in (\ref{PDG2012}) or Tables \ref{NormalH}, \ref{InvertedH}.
     This problem has been improved in Ref. \citen{dlnvS3} by adding a new triplet $\rho$
     and another antisextet $s'$, in which $s'$ is regarded as a small perturbation.
     Therefore, the model contains up to eight Higgs multiplets, and the scalar potential
of the model is quite complicated. In Ref. \citen{vlD4} we have studied the 3-3-1 model
with neutral fermions based
    on $D_4$ group, in which the fermion fields are in singlets and doublets under $D_4$.
Our aim in  this paper is to construct the 3-3-1 model combined
with $A_4$ to adapt non-zero $\theta_{13}$. For this purpose a
$SU(3)_L$ triplet is added and the result follows
without perturbation. We will work on a basis where $\underline{3}$ is a real representation.

There are two typical variants of the 3-3-1 models as far as
lepton sectors are concerned. In the minimal version, three
$\mathrm{SU}(3)_L$ lepton triplets are $(\nu_L,l_L,l^c_R)$, where
$l_{R}$ are ordinary right-handed charged leptons \cite{331m1} .
 In the second version, the third components of lepton triplets
are the  right-handed neutrinos\cite{331r1} , $(\nu_L,l_L,\nu^c_R)$. To have a model with the
realistic neutrino mixing matrix, we should consider another
variant of the form $(\nu_L,l_L,N^c_R)$ where $N_R$ are three new
fermion singlets under SM symmetry with vanishing
lepton-numbers \cite{dlshA4,dlsvS4} .

The contents of the paper are as follows. In Sec.
\ref{fermion} and Sec. \ref{clep} we present the necessary elements of the 3-3-1 model
with $A_4$ flavor symmetry as in the above choice and introduce necessary Higgs fields responsible for the
charged-lepton masses.
Sec. \ref{neutrino} is devoted for the neutrino masses and
mixings. In Sec. \ref{quarkA4}, we discuss the quark sector. We summarize our results and make conclusions in section \ref{conclus}. \ref{apa} presents a brief summary of the
$A_4$ group. \ref{apb} provides the lepton number ($L$)
and lepton parity ($P_l$) of  the particles in the model. Appendices from C to J
give the detailed solutions corresponding to special cases in the normal and inverted spectrum.
\section{Fermion content\label{fermion}}

The gauge symmetry is based on $\mathrm{SU}(3)_C\otimes
\mathrm{SU}(3)_L \otimes \mathrm{U}(1)_X$, where the electroweak
factor $\mathrm{SU}(3)_L \otimes \mathrm{U}(1)_X$ is extended from
those of the SM while the strong interaction sector is retained.
Each lepton family includes a new electrically- and
leptonically-neutral fermion $(N_R)$ and is arranged under the
$\mathrm{SU}(3)_L$ symmetry as a triplet $(\nu_L, l_L, N^c_R)$ and
a singlet $l_R$. The residual electric charge operator $Q$ is
related to the generators of the gauge symmetry by
\bea Q=T_3-\fr{1}{\sqrt{3}}T_8+X, \eea where $T_a$ $(a=1,2,...,8)$ are
$\mathrm{SU}(3)_L$ charges with $\mathrm{Tr}T_aT_b=\fr 1 2
\de_{ab}$ and $X$ is the  $\mathrm{U}(1)_X$ charge. The model
under consideration does not contain exotic electric charges in
the fundamental fermion, scalar and adjoint gauge boson
representations.

Since the particles in the lepton triplet have different lepton
number (1 and 0), so the lepton number in the model  does not
commute with the gauge symmetry unlike the SM. Therefore, it is
better to work with a new conserved charge $\mathcal{L}$\cite{changlong,Tully, Buras,Boucenna} commuting
with the gauge symmetry and related to the ordinary lepton number
by diagonal matrices \cite{dlshA4,dlsvS4}
 \bea L=\fr{2}{\sqrt{3}}T_8+\mathcal{L}.\eea
The lepton charge arranged in this way (i.e. $L(N_R)=0$ as
assumed) is in order to prevent unwanted interactions due to
$\mathrm{U}(1)_\mathcal{L}$ symmetry and breaking to obtain the consistent lepton and
quark spectra. By this embedding, exotic quarks $U, D$ as well as
new non-Hermitian gauge bosons $X^0$, $Y^\pm$ possess lepton
charges as of the ordinary leptons:
$L(D)=-L(U)=L(X^0)=L(Y^{-})=1$.

In the model under consideration, the fermion contents is same as in \cite{dlshA4} . However,
 this work is distinguished by a new $SU(3)_L$ triplet ($\rho$) which is put in $\underline{3}$ under $A_4$.
 Under the $[\mathrm{SU}(3)_L, \mathrm{U}(1)_X,
\mathrm{U}(1)_\mathcal{L},\underline{A}_4]$ symmetries, the fermions of the model
 transform as follows \cite{dlshA4}
\bea \psi_{L} &\equiv& \psi_{1,2,3L}=\left( \nu_{L} \,\, l_{L}
\,\,  N^c_{R}\right)^T\sim [3,-1/3,2/3,\underline{3}],\crn
l_{1R}&\sim&[1,-1,1,\underline{1}],\hs
l_{2 R}\sim[1,-1,1,\underline{1}'], \hs
l_{3 R}\sim [1,-1,1,\underline{1}''],\label{Clepc}\\
Q_{3L}&=&
\left(%
\begin{array}{c}
  u_{3L} \\
  d_{3L} \\
  U_L \\
\end{array}%
\right)\sim [3,1/3,-1/3,\underline{1}],\hs U_R \sim
[1,2/3,-1,\underline{1}],\crn
Q_{1L}&=&
\left(%
\begin{array}{c}
  d_{1L} \\
  -u_{1L} \\
  D_{1L} \\
\end{array}%
\right)\sim [3^*,0,1/3,\underline{1}'],\hs D_{1R}\sim
[1,-1/3,1,\underline{1}''], \crn
Q_{2L}&=&
\left(%
\begin{array}{c}
  d_{2L} \\
  -u_{2L} \\
  D_{2L} \\
\end{array}%
\right)\sim [3^*,0,1/3,\underline{1}''],\hs D_{2R}\sim
[1,-1/3,1,\underline{1}'],\crn
u_R &\sim&
[1,2/3,0,\underline{3}],\hs d_R \sim
[1,-1/3,0,\underline{3}],\label{quark}
\eea where the subscript
numbers on field indicate to respective families which also define
components of their $A_4$ multiplets. In what follows, we consider
possibilities of generating the masses for the fermions. The
scalar multiplets needed for the purpose are also introduced.

\section{Charged lepton mass \label{clep}}
The fermion content of the model is the same as that
 in  Ref.\citen{dlshA4} under all symmetries. However, in this work the breaking of $A_4$ in charged lepton sector is different from that
 in Ref.\citen{dlshA4} . Namely, to generate
masses for the charged leptons, we need only one scalar
multiplet:
\bea \phi = \left(%
  \phi^+_1, \,
  \phi^0_2,  \,
  \phi^+_3
\right)^T \sim [3,2/3,-1/3, \underline{3}]. \label{phi}\eea
The Yukawa terms are \bea -\mathcal{L}_{l}&=&h_1 (\bar{\psi}_{L}
\phi)_{\underline{1}} l_{1R}+h_2 (\bar{\psi}_{L}
\phi)_{\underline{1}''} l_{2 R} +h_3 (\bar{\psi}_{L}
\phi)_{\underline{1}'} l_{3 R}+H.c.\label{Yclep} \eea
From
the potential minimization conditions, we have the followings
alignments: \bit
\item[(1)] The first alignment: $\langle \phi_1\rangle= \langle \phi_2\rangle=\langle \phi_3\rangle$ then $A_4$
 is broken into $Z_3$ consisting of the elements \{$e, T, T^2$\}.
\item[(2)] The second alignment: $\langle \phi_1\rangle\neq \langle \phi_2\rangle\neq\langle \phi_3\rangle$
or $\langle \phi_1\rangle\neq \langle \phi_2\rangle=\langle
\phi_3\rangle$ or $\langle \phi_2\rangle\neq \langle
\phi_1\rangle=\langle \phi_3\rangle$ or $\langle \phi_3\rangle\neq
\langle \phi_1 \rangle=\langle \phi_2\rangle$ then $A_4$ is broken
into $\{\mathrm{Identity}\}$\footnote{This means $A_4$ is completely broken.}.
\item[(3)] The third alignment: $0=\langle \phi_1\rangle\neq\langle \phi_2\rangle=\langle \phi_3\rangle
\neq 0$ or $0=\langle \phi_2\rangle\neq\langle
\phi_3\rangle=\langle \phi_1\rangle \neq 0$ or $0=\langle
\phi_3\rangle\neq\langle \phi_1\rangle=\langle \phi_2\rangle\neq
0$ then $A_4$ is broken into $\{\mathrm{Identity}\}$.
\item[(4)] The fourth alignment: $0=\langle \phi_1\rangle\neq\langle \phi_2\rangle\neq\langle
\phi_3\rangle\neq 0$ or $0= \langle \phi_2\rangle\neq\langle
\phi_1\rangle\neq\langle \phi_3 \rangle\neq 0$ or $0=\langle
\phi_3\rangle\neq\langle \phi_1\rangle\neq\langle \phi_2\rangle
\neq0$ then $A_4$ is broken into $\{\mathrm{Identity}\}$.
\item[(5)] The fifth alignment: $0=\langle \phi_2\rangle=\langle \phi_3\rangle\neq\langle
\phi_1\rangle\neq0$ then $A_4$ is broken into $Z_2$ consisting of
the elements \{$e, S$\}.
\item[(6)] The sixth alignment: $0=\langle \phi_1\rangle=\langle \phi_3\rangle\neq\langle
\phi_2\rangle\neq0$ then $A_4$ is broken into $Z_2$ consisting of
the elements \{$e, T^2ST$\}.
\item[(7)] The seventh alignment: $0=\langle \phi_1\rangle=\langle \phi_2\rangle\neq\langle \phi_3
\rangle\neq0$ then $A_4$ is broken into $Z_2$ consisting of the
elements \{$e, TST^2$\}. \eit
To obtain a realistic lepton spectrum, we suppose that in
charged lepton sector $A_4$ is broken down to
$\{\mathrm{Identity}\}$. This breaking is different from Ref. \citen{dlshA4} in charged lepton sector,
and it can be achieved with the VEV
alignment $\langle \phi\rangle=(\langle \phi_1\rangle, \langle \phi_2 \rangle, \langle \phi_3\rangle)$
under $A_4$ where $\langle \phi_1\rangle\neq \langle \phi_2
\rangle\neq\langle \phi_3\rangle$, and
  \[  \langle\phi_i\rangle =\left(  0 \,\,\,\,\,  v_i \,\,\,\,\,  0\right)^T,\,\,\, (i=1,2,3).\]
  The mass
Lagrangian for the charged leptons reads \[
\mathcal{L}^{\mathrm{mass}}_l=-(\bar{l}_{1L},\bar{l}_{2L},\bar{l}_{3L})
M_l (l_{1R},l_{2R},l_{3R})^T+H.c,\]
where \be M_l=
\left(%
\begin{array}{ccc}
  h_1v_1 & h_2v_1 & h_3 v_1 \\
   h_1v_2 & h_2\om v_2 & \,\, h_3\om^2 v_2 \\
  h_1v_3 &\,\,\, h_2\om^2 v_3 &\,\,h_3\om v_3 \\
\end{array}%
\right).\label{Mlep}\ee As will see in section \ref{neutrino}, in
the case $A_4\rightarrow Z_3$ consisting of the elements \{$e, T,
T^2$\}, i.e, $\langle \phi_1\rangle =\langle\phi_2 \rangle=\langle
\phi_3\rangle$ or $v_1=v_2=v_3=v$, the charged lepton matrix $M_l$ in Eq. (\ref{Mlep})
is diagonalized by the matrix
\be U_{0L}=\fr{1}{\sqrt{3}}\left(%
\begin{array}{ccc}
  1 & 1 & 1 \\
  1 & \om & \om^2 \\
  1 & \om^2 & \om \\
\end{array}%
\right),\label{U0clep}\ee
and the exact tri-bimaximal mixing
form is obtained if $A_4\rightarrow Z_3$ in both charged lepton
and neutrino sectors. A detail study on this problem, the reader can see in Ref. \citen{dlshA4}.

As we know, the realistic lepton mixing form is a small deviation
from tri-bimaximal form \cite{PDG2012} . The realistic lepton mixing can be achieved with a small value.
 Hence, we can separate $v_2, v_3$ into two parts, the first is equal to $v_1\equiv v$,
 the second is responsible for that deviation, \bea v_1&=&v, \,\,
v_2=v (1+\varep_2), \,\, v_3=v (1+\varep_3), \,\, \varep_{2,3}\ll
1,\label{epsi12} \eea
and the matrix $M_l$ in (\ref{Mlep}) becomes
\bea M_l&=&
\left(%
\begin{array}{ccc}
  h_1v & h_2v & h_3 v \\
   h_1v(1+\varep_2) & h_2\om v(1+\varep_2) & \,\, h_3\om^2 v(1+\varep_2) \\
  h_1v(1+\varep_3) &\,\,\, h_2\om^2 v(1+\varep_3) &\,\,h_3\om v(1+\varep_3) \\
\end{array}%
\right)\crn &\equiv&v\left(%
\begin{array}{ccc}
  1 & 0 & 0 \\
  0 &\,\,\, 1+\varep_2 & 0 \\
  0 &0 &1+\varep_3 \\
\end{array}%
\right)\left(%
\begin{array}{ccc}
  1 & 1 & 1 \\
  1 & \om & \om^2 \\
  1 & \om^2 & \om \\
\end{array}%
\right)\left(%
\begin{array}{ccc}
  h_1 & 0& 0 \\
  0 & h_2 & 0\\
  0 &0&h_3 \\
\end{array}%
\right).\label{Mlep1}\eea
The matrix $M_l$ in Eq. (\ref{Mlep1})
can be diagonalized as follows:\\
Denoting
 \bea
&&M'_l=U^+_{0L}
M_l\label{Mlepp}=\frac{v}{\sqrt{3}}\left(%
\begin{array}{ccc}
  (3+\varep_1+\varep_2)h_1 & (\om \varep_1+\om^2 \varep_2)h_2& (\om^2 \varep_1+\om \varep_2)h_3 \\
  (\om^2 \varep_1+\om \varep_2)h_1 & (3+\varep_1+\varep_2)h_2 & (\om\varep_1+\om^2\varep_2)h_3\\
  \om\varep_1+\om^2\varep_2)h_1 &(\om^2\varep_1+\om\varep_2)h_2&(3+\varep_1+\varep_2)h_3 \\
\end{array}%
\right),\label{Mlepp}\eea
then the matrix $M'_l$ in (\ref{Mlepp})
is diagonalized by
 \be U^+_L M'_l \equiv  U^+_L U^+_{0L}M_l=
\mathrm{diag} (m_e, m_\mu , m_\tau),
\label{Mleppv}\ee
 where
 \be
 m_e=Y_l h_1 v,\,\,
m_\mu= Y_l h_2 v,\,\, m_\tau=Y_l h_3 v, \label{memutau}
 \ee
 with
 \bea
 Y_l&=&\frac{3\sqrt{3}(1 + \varep_3)[-4+
        \varep_3(-4 +
\varep_3+ \sqrt{(\varep_3-12)\varep_3-12})]}{(2+\varep_3)[-6+
        \varep_3(-6 +
\varep_3+ \sqrt{(\varep_3-12)\varep_3-12})]}.\label{Y}
 \eea
 The matrix that diagonalize $M'_l$ in (\ref{Mlepp}) takes the form:
  \be
U_L=\left(%
\begin{array}{ccc}
1&U^l_{12}&U^l_{13} \\
U^l_{13}&1&U^l_{12} \\
U^l_{12}&U^l_{13}&1 \\
\end{array}%
\right),\hs U_R =1 \label{Uclep} \ee 
where
\bea U^l_{12}&=&\frac{\varep_3\left\{6-2i\sqrt{3}-(1+i\sqrt{3})\varep+\varep_3[7-i\sqrt{3}-(1-i\sqrt{3})\varep_3+(1-i\sqrt{3})\varep]\right\}}{2(2+\varep_3)[-6+\varep^2_3-\varep_3(6+\varep)]},\crn
U^l_{13}&=&\frac{\varep_3\left\{6+2i\sqrt{3}-(1-i\sqrt{3})\varep+\varep_3[7+i\sqrt{3}-(1+i\sqrt{3})\varep_3+(1+i\sqrt{3})\varep]\right\}}{2(2+\varep_3)[-6+\varep^2_3-\varep_3(6+\varep)]},\hs\label{u12u13l}\eea
with
\bea
\varep&=&\sqrt{\varep^2_3-12(\varep_3+1)}.\label{varep}\eea
 To get the results in
Eqs.(\ref{u12u13l}) we have used the following relations \bea
\varep_2&=&\frac{2\varep_3 -\varep^2_3 -\varep_3\varep}{2(\varep_3+2)},\hs
\varep^*_2=\frac{\varep_3\left(-2-3\varep_3 +\varep\right)}{2(\varep_3+1)(\varep_3+2)}, \crn
\varep^*_3&=&-1+\frac{1}{1 + \varep_3},\label{eeprelat}
 \eea
 which are obtained from the unitary condition of $U_L$.\\
 The left- and right- handed mixing matrices in charged lepton sector are given by:
 \be
 U'_L=U_{0L}.U_{L}=\left(%
\begin{array}{ccc}
\al_1&\al_2&\al_1 \\
\al_2&\om^2 \al_2&\om \al_2 \\
\al_3&\om \al_3&\om^2 \al_3 \\
\end{array}%
\right),\,\, U'_R=1, \label{Uclepp}\ee
where
\bea
\al_1&=&\frac{\sqrt{3}\left[-4+\varep^2_3-\varep_3 (4+\varep)\right]}{(2+\varep_3)[-6+\varep^2_3-\varep_3(6+\varep)]},\crn
\al_2&=&\frac{2\sqrt{3}(1+\varep_3)}{6-\varep^2_3+\varep_3(6+\varep)},\hs
\al_3=(1+\varep_3)\al_1 . \label{al123}
\eea
In general, $\varep_{2,3} \neq 0$, so $\al_i\,\, (i=1,2,3)$ in Eq. (\ref{al123}) are different to
each other and different from $\frac{1}{\sqrt{3}}$, and lead to the realistic lepton mixing with
non-zero $\theta_{13}$ as represented in Sec.\ref{neutrino}. This is one of the striking results of the model under consideration.

Taking into account of the discovery of the long-awaited Higgs boson at
 around 125 GeV by ATLAS\cite{ATLAS}  and
CMS \cite{CMS} , we can choose
the VEVs $v=100\,  \rm{GeV}$. From (\ref{memutau}), the charged lepton Yukawa
couplings $h_{1,2,3}$ relate to their masses as follows:
\bea
h_1&=&\frac{m_e}{Y_l v},\,\, h_2=\frac{m_\mu}{Y_l v},\,\, h_3=\frac{m_\tau}{Y_l v}.\label{h123}
\eea
The experimental mass values for the charged leptons at the weak
scale are given as \cite{PDG2012} : \bea m_e&\simeq&0.511\, \textrm{MeV},\,\,
m_{\mu}\simeq 105.66\, \textrm{MeV},\,\,
m_{\tau}\simeq1776.82\, \textrm{GeV}\label{memutaexp} \eea
With the help of (\ref{memutaexp}) we have $\frac{h_1}{h_2}\simeq0.0048, \,\, \frac{h_1}{h_3}\simeq 0.0003$ and $\frac{h_2}{h_3}=0.0595$, i.e, $h_1\ll h_2\ll h_3$ for any $\varep_3$.
As will be shown in Sec.\ref{neutrino}, from experimental constrains on lepton mixing, we obtain two solutions in Eqs. (\ref{ep31}) and (\ref{ep32}). With $\varep_3$ given in Eq.(\ref{ep31}), we get
\bea h_1&\simeq&3.0045\times 10^{-6}, \,\,
h_2\simeq6.2124\times 10^{-4}, \,\,
h_3\simeq 1.045\times 10^{-2}.\label{h1h2h3}\eea
We note that the mass hierarchy of the charged leptons are well separated by only one Higgs triplet $\phi$, and this is a good feature
 of the $A_4$ group. To conclude this section, we remind that the situation here is
different from all our previous version presented in Refs.\citen{dlshA4,dlsvS4,dlnvS3,vlD4}
that can lead to non-zero $\theta_{13}$ which is studied in section \ref{neutrino}.
 \section{\label{neutrino} Neutrino mass and mixing}

The neutrino masses arise from the couplings of $\bar{\psi}^c_{L} \psi_{L}$ to scalars,
 where $\bar{\psi}^c_{L} \psi_{L}$ transforms as
$3^*\oplus 6$ under $\mathrm{SU}(3)_L$ and $\underline{1}\oplus \underline{1}'\oplus \underline{1}''\oplus
\underline{3}_s\oplus
\underline{3}_a$ under $A_4$. For the known scalar triplets, there is no interactions invariant
 under all subgroups of $G=\mathrm{SU}(3)_C\otimes
\mathrm{SU}(3)_L \otimes \mathrm{U}(1)_X \otimes A_4$. We
will therefore propose new SU(3)$_L$ antisextets, lying in either $\underline{1}$, $\underline{1}'$,
$\underline{1}''$, or $\underline{3}$ under $A_4$ interacting with $\bar{\psi}^c_{L}\psi_{L}$
 to produce masses for the neutrinos. Therefore, new $SU(3)_L$ anti-sextets are proposed.
 The antisextets transform as follows:
\bea \sigma&=&
\left(%
\begin{array}{ccc}
  \sigma^0_{11} & \sigma^+_{12} & \sigma^0_{13} \\
  \sigma^+_{12} & \sigma^{++}_{22} & \sigma^+_{23} \\
  \sigma^0_{13} & \sigma^+_{23} & \sigma^0_{33} \\
\end{array}%
\right)\sim \left[6^*,\frac{2}{3},-\frac{4}{3}, \underline{1}\right], \label{sis}\\
s_i&=&
\left(%
\begin{array}{ccc}
  s^0_{11} & s^+_{12} & s^0_{13} \\
  s^+_{12} & s^{++}_{22} & s^+_{23} \\
  s^0_{13} & s^+_{23} & s^0_{33} \\
\end{array}%
\right)_i\sim \left[6^*,\frac{2}{3},-\frac{4}{3}, \underline{3}\right], \, (i=1,2,3). \nn\eea

Following the potential minimization
conditions, we have the followings alignments:
\begin{itemize}
\item[(1)] The first alignment: $\langle s_1\rangle= \langle s_2\rangle=\langle s_3\rangle$ then $A_4$
 is broken into $Z_3$ consisting of the elements \{$e, T, T^2$\}.

\item[(2)] The second alignment: $\langle s_1\rangle\neq \langle s_2\rangle\neq\langle s_3\rangle$
or $\langle s_1\rangle\neq \langle s_2\rangle=\langle s_3\rangle$ or $\langle s_2\rangle\neq
\langle s_1\rangle=\langle s_3\rangle$ or $\langle s_3\rangle\neq \langle s_1
\rangle=\langle s_2\rangle$ then $A_4$ is broken into $\{\mathrm{Identity}\}$.

\item[(3)] The third alignment: $0=\langle s_1\rangle\neq\langle s_2\rangle=\langle s_3\rangle
\neq 0$ or $0=\langle s_2\rangle\neq\langle s_3\rangle=\langle s_1\rangle \neq 0$
or $0=\langle s_3\rangle\neq\langle s_1\rangle=\langle s_2\rangle\neq 0$
then $A_4$ is broken into $\{\mathrm{Identity}\}$.

\item[(4)] The fourth alignment: $0=\langle s_1\rangle\neq\langle s_2\rangle\neq\langle
s_3\rangle\neq 0$ or $0= \langle s_2\rangle\neq\langle s_1\rangle\neq\langle s_3
\rangle\neq 0$ or $0=\langle s_3\rangle\neq\langle s_1\rangle\neq\langle s_2\rangle
\neq0$  then $A_4$ is broken into $\{\mathrm{Identity}\}$.

\item[(5)] The fifth alignment: $0=\langle s_2\rangle=\langle s_3\rangle\neq\langle
s_1\rangle\neq0$ then $A_4$ is broken into $Z_2$ consisting of the elements \{$e, S$\}.

\item[(6)] The sixth  alignment: $0=\langle s_1\rangle=\langle s_3\rangle\neq\langle
s_2\rangle\neq0$ then $A_4$ is broken into $Z_2$ consisting of the elements \{$e, T^2ST$\}.

    \item[(7)] The seventh alignment: $0=\langle s_1\rangle=\langle s_2\rangle\neq\langle s_3
\rangle\neq0$ then $A_4$ is broken into $Z_2$ consisting of the elements \{$e, TST^2$\}.
\end{itemize}
To obtain a realistic neutrino spectrum, we
argue that the breaking $A_4\rightarrow Z_2$ must be taken place. This can be achieved
within each case below.
\begin{itemize}
\item A new $SU(3)_L$ anti-sextet $s$ given in (\ref{sis}),  with the VEVs chosen by $\langle s\rangle=(\langle
 s_1\rangle, 0, 0)$ under $A_4$, where
\be \langle s_1\rangle=\left(%
\begin{array}{ccc}
  \la_{s} & 0 & v_{s} \\
  0 & 0 & 0 \\
  v_{s} & 0 & \La_{s} \\
\end{array}%
\right). \label{svev} \ee

\item  Another  $\mathrm{SU}(3)_L$ triplet $\rho$ which is also put in the $\underline{3}$ under $A_4$:
\[ \rho_i = \left(%
  \rho^+_1, \,
  \rho^0_2 , \,
  \rho^+_3
\right)^T_i\sim [3,2/3,-4/3, \underline{3}], \, (i=1,2,3) \nn\]
with the VEV chosen by
 \bea
 \langle\rho \rangle=\left(\langle
 \rho_1\rangle, 0, 0\right),\hs \langle \rho_1 \rangle = (0,\,\,v_\rho,\,\,0)^T.\label{vevrho}\eea
\end{itemize}

In this work, we additionally introduce  a new $SU(3)_L$ triplet $\rho$ lying in $\underline{3}$ under $A_4$
to obtain non-zero $\theta_{13}$,  which is different from that  in Refs. \citen{dlshA4,dlsvS4} .

The neutrino Yukawa interactions are 
\bea -\mathcal{L}_\nu&=&\fr x 2(\bar{\psi}^c_{L}\psi_{L})_{\underline{1}}\sigma
+\fr y 2(\bar{\psi}^c_{L}\psi_{L})_{\underline{3}}s
+\fr z 2(\bar{\psi}^c_{L}\psi_{L})_{\underline{3}}\rho+H.c\crn
&=&\fr x 2(\bar{\psi}^c_{1L}\psi_{1L}+\bar{\psi}^c_{2L}\psi_{2L}+\bar{\psi}^c_{3L}\psi_{3L})\sigma\crn
&+&y(\bar{\psi}^c_{2L}\psi_{3L}s_1+\bar{\psi}^c_{3L}\psi_{1L}s_2+\bar{\psi}^c_{1L}\psi_{2L}s_3)\crn
&+&\fr z 2 \left[(\bar{\psi}^c_{ 2L} \rho_{3}-\bar{\psi}^c_{3L}\rho_{2})\psi_{1L}+
(\bar{\psi}^c_{ 3L} \rho_{1}-\bar{\psi}^c_{1L}\rho_{3})\psi_{2L}\right.\crn
&+&\left.(\bar{\psi}^c_{1L} \rho_{2}-\bar{\psi}^c_{2L}\rho_{1})\psi_{3L}\right]
+H.c.\label{Lnu0}\eea
{\rm With}  the VEV of $\sigma$ is \be
\langle \sigma \rangle=\left(%
\begin{array}{ccc}
  \la_{\si} & 0 & v_{\si} \\
  0 & 0 & 0 \\
  v_{\si} & 0 & \La_{\si} \\
\end{array}%
\right),\label{vevsi}\ee
the mass Lagrangian for the neutrinos can be written in matrix form:
 \be -\mathcal{L}^{\mathrm{mass}}_\nu=\fr 1 2
\bar{\chi}^c_L M_\nu \chi_L+ H.c.,\label{nm}\ee where
\bea \chi_L&\equiv&
\left(\nu_L \hs
  N^c_R \right)^T,\,\,\,\,\,\, M_\nu\equiv\left(%
\begin{array}{cc}
  M_L & M^T_D \\
  M_D & M_R \\
\end{array}%
\right), \label{MLDR}\\
\nu_L&=&(\nu_{1L},\nu_{2L},\nu_{3L})^T,\,
N_R=(N_{1R},N_{2R},N_{3R})^T,\nn \eea
and the mass matrices are then obtained by
\be
M_{L,R,D}=\left(%
\begin{array}{ccc}
  a_{L,R,D} & 0 & 0 \\
  0 & a_{L,R,D} &b_{L,R,D}+d_{L,R,D} \\
 0 & b_{L,R,D}-d_{L,R,D} &a_{L,R,D} \\
\end{array}%
\right),\label{abcd}\ee
with
\bea
  a_{L} & =&\la_\si x, \hs  a_{D} =v_\si x, \hs  a_{R} =\La_\si x, \crn
  b_{L} & =&\la_sy ,\hs  b_{D}=v_sy,\hs  b_{R} =\La_sy,\crn
  d_{L} & =&d_{R} =0, \hs  d_{D} =v_\rho z. \label{abdLDR}\eea
Three observed neutrinos gain masses via a combination of type I
and type II seesaw mechanisms derived from (\ref{MLDR}) and (\ref{abcd}) as \be
M_{\mathrm{eff}}=M_L-M_D^TM_R^{-1}M_D=\left(%
\begin{array}{ccc}
  A & 0 & 0 \\
  0 & B_1 & C \\
  0 & C & B_2 \\
\end{array}%
\right), \label{Mef}\ee where \bea
A&=& a_L-\frac{a^2_D}{a_R},\crn
B_1&=&a_L-\frac{a^2_D a_R+a_R(b_D-d_D)^2-2a_Db_R(b_D-d_D)}{a^2_R-b^2_R}\crn
B_2&=&B_1+\frac{4 (a_D b_R-a_R b_D )d_D}{a^2_R-b^2_R},\crn
C&=&b_L+\frac{b_R(a^2_D+b^2_D-d^2_D)-2 a_D a_R b_D}{a^2_R-b^2_R}.\label{ABC}
\eea
We can diagonalize the mass matrix (\ref{Mef}) as follows
\[ U^T_\nu
 M_{\mathrm{eff}} U_\nu=\mathrm{diag}(m_1,  m_2,  m_3),\] with
\bea m_1
&=&\fr 1 2 \left(B_1 + B_2 + \sqrt{(B_1 -B_2)^2+4C^2}\right),\crn
m_2&=&A,\label{m123}\\
m_3&=&\fr 1 2 \left(B_1 + B_2 - \sqrt{(B_1- B_2)^2+4C^2}\right),\nn\eea
and the corresponding neutrino mixing matrix: \be U_\nu=\left(%
\begin{array}{ccc}
  0 & 1 & 0 \\
  \fr{1}{\sqrt{K^2+1}} & 0 & \fr{K}{\sqrt{K^2+1}} \\
  -\fr{K}{\sqrt{K^2+1}} & 0 & \fr{1}{\sqrt{K^2+1}} \\
\end{array}%
\right)\times P,\label{Unu1}\ee
where $P=\mathrm{diag}(1,1,i)$, and 
\be
K=\frac{B_1 -B_2 -\sqrt{(B_1-B_2)^2+4 C^2}}{2C}.\label{K}
\ee
Note that $K$ in Eq.(\ref{K}) must be a real number since the unitary condition of $U_\nu$. 
Combined with (\ref{Uclepp}) and (\ref{Unu1}), the lepton mixing matrix yields the form:
 \bea
U_{lep}=U'^{+}_L U_\nu=\left(%
\begin{array}{ccc}
U_{11}&U_{12}&U_{13} \\
U_{21}&U_{22}&U_{23} \\
U_{31}&U_{32}&U_{33} \\
\end{array}%
\right)\times P,\label{Umpns} \eea where \bea U_{11}&=&-\frac{\sqrt{3}\left\{4(1-K)+\varep_3\left[6+(\varep-4)K+\varep_3(K+2)\right]\right\}}
{(2+\varep_3)\left[-6+\varep_3(-6+\varep_3+\varep)\right]\sqrt{K^2+1}},\crn
U_{12}&=&U_{22}=U_{32}=\frac{\sqrt{3}(1+\varep_3)\left[-4+\varep_3(-4+\varep_3+\varep)\right]}
{(2+\varep_3)\left[-6+\varep_3(-6+\varep_3+\varep)\right]},\crn
U_{13}&=&-\frac{\sqrt{3}\left\{4(1+K)+\varep_3\left[4-\varep+6K+\varep_3(2K-1)\right]\right\}}
{(2+\varep_3)\left[-6+\varep_3(-6+\varep_3+\varep)\right]\sqrt{K^2+1}},\crn
U_{21}&=&\frac{2(-3i+\sqrt{3})(1+\varep_3)+\frac{(3i+\sqrt{3})\left[-4+\varep_3(-4+\varep_3+\varep)\right]K}{2+\varep_3}}
{2\left[-6+\varep_3(-6+\varep_3+\varep)\right]\sqrt{K^2+1}},\crn
U_{23}&=&-\frac{\frac{(3i+\sqrt{3})\left[-4+\varep_3(-4+\varep_3+\varep)\right]}{2+\varep_3}-2(-3i+\sqrt{3})(1+\varep_3)K}
{2\left[-6+\varep_3(-6+\varep_3+\varep)\right]\sqrt{K^2+1}},\crn
U_{31}&=&\frac{2(3i+\sqrt{3})(1+\varep_3)+\frac{(-3i+\sqrt{3})\left[-4+\varep_3(-4+\varep_3+\varep)\right]K}{2+\varep_3}}
{2\left[-6+\varep_3(-6+\varep_3+\varep)\right]\sqrt{K^2+1}},\crn
U_{33}&=&-\frac{\frac{(-3i+\sqrt{3})\left[-4+\varep_3(-4+\varep_3+\varep)\right]}{2+\varep_3}-2(3i+\sqrt{3})(1+\varep_3)K}
{2\left[-6+\varep_3(-6+\varep_3+\varep)\right]\sqrt{K^2+1}}, \label{Ulij}\eea
with $\varep$ is defined in Eq.(\ref{varep}).
We see that all the elements of the matrix $U_{lep}$ in Eq. (\ref{Ulij}) depend only on
 one parameter $\varep_3$.
From experimental constraints on the elements of the lepton mixing matrix given in Eq.(\ref{Uij}), we can find out the regions of $K$ and $\varep_3$ that satisfy experimental data on lepton mixing matrix.
The good value of $K$ is in one of the following regions:
\bea
&&K\in (-1.45,-1.4), \,\,\, K\in (1.4,1.45),\crn
&& K\in (-0.75,-0.65), \,\,\, K\in (0.65,0.75).\label{Kregion}\eea
At present the values of the
absolute neutrino masses as well as the mass ordering of
neutrinos are still open problems. An upper bound
on the absolute value of neutrino mass was found from the analysis
of the latest cosmological data \cite{Tegmark} \be m_i\leq 0.6\,
\mathrm{eV},\label{upb} \ee
The 95\% upper limit on the sum of neutrino mass is given in Ref. \citen{planck}
\be \sum^{3}_{i=1} |m_i|\leq 0.66\,  \mathrm{eV}.
\label{upbsum}\ee
The mass
ordering of neutrino depends on the sign of $\Delta m^2_{13}$
which is currently unknown. In the case of 3-neutrino mixing, the two possible signs of $\Delta
m^2_{13}$ corresponding to two types of neutrino mass spectrum can
be provided as follows

\ben \item Normal hierarchy (NH): $
|m_1|\simeq |m_2| < |m_3|,\,\, \Delta m^2_{31}=m^2_3-m^2_1>0.$
\item Inverted hierarchy (IH): $|m_3|< |m_1|\simeq |m_2|,\,\, \Delta m^2_{31} =m^2_3-m^2_1<0$.\een
As will be discussed below, the neutrino mass  matrix  in (\ref{Mef}) can provide both normal and inverted
 mass hierarchies.

In this work, to have explicit values of the model parameters, the values of $K$:\, $K=-1.43, \, K=1.43, \, K=-0.7$ and $K=0.7$ [which all satisfy (\ref{Kregion})] are used. The corresponding expressions of $B_{1,2}, C$,
and $m_{1,2,3}$ are given in Appendices {\rm from} \ref{K1N} to \ref{K4N}  for normal hierarchy
 and in Appendices {\rm from} \ref{K1I} to \ref{K4I} for inverted hierarchy. However,
 the corresponding physical results  such as the values of the
absolute neutrino masses are the same. So, here we only consider in detail the
 case $K=-1.43$ for both normal and inverted spectrum.

Combining with the constraint values on the element $U_{11}$ of lepton mixing
 matrix \cite{Uij} , $U_{11}=0.812$, we obtain two {\rm solutions} on  $\varep_3$:
\be
\varep_3=-0.0318467 - 0.00695743 i, \label{ep31}
\ee
and
\be
\varep_3=-0.0318467 + 0.00695743i. \label{ep32}
\ee
With the solution (\ref{ep31}), it  follows:
\bea
U_{lep}&\simeq&\left(%
\begin{array}{ccc}
 0.812                 & 0.567+ 0.003i& -0.139 + 0.013i \\
 -0.395 - 0.128i & 0.567+ 0.003i & 0.067 - 0.708i\\
-0.417 + 0.128i  & 0.567+ 0.003i &0.074+ 0.695i \\
\end{array}%
\right)\times P, \label{Ulep1}\eea
or
\be
|U_{lep}|=\left(%
\begin{array}{ccc}
0.812 &\hs 0.567&\hs 0.140 \\
0.415&\hs 0.567&\hs 0.711\\
0.436&\hs 0.567&\hs 0.699 \\
\end{array}%
\right), \label{Ulepa1} \ee
and
$\varep_2=-0.0224354 + 0.0238703 i$.

With the solution (\ref{ep32}), we get:
\bea
U_{lep}&=&\left(%
\begin{array}{ccc}
 0.812                & 0.567- 0.003i& -0.139- 0.013i   \\
 -0.417 - 0.128i & 0.567- 0.003i& 0.074 - 0.695i\\
 -0.395+ 0.128i& 0.567- 0.003i& 0.067 + 0.708 i \\
\end{array}%
\right)\times P, \label{Ulep2}\eea
or
\be
|U_{lep}|=\left(%
\begin{array}{ccc}
0.812 &\hs 0.567&\hs 0.140 \\
0.436&\hs 0.567&\hs 0.699 \\
0.415&\hs 0.567&\hs 0.711\\
\end{array}%
\right), \label{Ulepa2} \ee
and
$\varep_2=-0.0224354 - 0.0238703 i$

In the standard Particle Data Group(PDG) parametrization, the lepton mixing
 matrix can be parametrized as
 \bea
U_{PMNS}&=&\left(%
\begin{array}{ccc}
 c_{12} c_{13}                                                       & -s_{12} c_{13} &  - s_{13} e^{-i \delta} \\
    s_{12} c_{23}-c_{12} s_{23} s_{13}e^{i \delta} & c_{12} c_{23}+s_{12} s_{23} s_{13} e^{i \delta}& - s_{23} c_{13}\\
    s_{12} s_{23}+c_{12} c_{23} s_{13}e^{i \delta}& c_{12} s_{23}-s_{12} c_{23} s_{13} e^{i \delta}& c_{23} c_{13} \\
\end{array}%
\right)\times \mathcal{P}, \label{Ulepg}\eea
where $\mathcal{P}=\mathrm{diag}(1, e^{i \alpha}, e^{i \beta})$, and
$c_{ij}=\cos \theta_{ij}$, $s_{ij}=\sin \theta_{ij}$ with
$\theta_{12}$, $\theta_{23}$ and $\theta_{13}$ being the
solar, atmospheric and  reactor angles, respectively.
$\delta= [0, 2\pi]$ is the Dirac CP violation phase while $\alpha$ and
$\beta$ are two Majorana CP violation phases.
The observable angles in the standard PMNS parametrization are given by \cite{PDG2012}
\bea
	s_{13} &=& \left| U_{13} \right|, \,\,
	s_{23} =  \frac{\left| U_{23} \right|}{\sqrt{1- \left| U_{1 3} \right|^2 }} , \,\,
	s_{12} = \frac{\left| U_{12} \right|}{\sqrt{1- \left| U_{13} \right|^2} }.\label{thetaij}
\eea
Combining Eq.(\ref{Ulep1}) and Eq. (\ref{thetaij}) yields:
\[
	\sin\theta_{13} =0.140,\,\, \sin \theta_{23} =0.719,\,\,	\sin \theta_{12} =0.573,\]
or
\[
\theta_{13} \simeq 8.055^{\circ},\,\, \theta_{23} \simeq 45.95^{\circ},\,\,	\theta_{12} \simeq 34.93^{\circ},\]
which are all very consistent with the recent data on neutrino mixing angles.
On the other hand, comparing Eq. (\ref{Ulij}) and (\ref{Umpns}) yields $\alpha=0, \beta =\frac{\pi}{2}$, and $\delta=5.41^{\circ}$ since $e^{i\delta}=-s_{13}/U_{13}=0.995547 + 0.0942709 i$.
These results also implies that in the
model under consideration, the value of the Jarlskog invariant
$J_{CP}$ which determines the magnitude of CP violation in
neutrino oscillations is determined \cite{Jarlskog} : \bea
J_{CP}&=& \frac{1}{8}\cos\theta_{13}\sin2\theta_{12}\sin2\theta_{23}\sin2
\theta_{13}\sin\delta
=0.003.\label{J1}
\eea
Similarly, with the solution (\ref{ep32}),
we get the results given in
 Tab. \ref{resultaN}.
\begin{table}[h]
\tbl{The model parameters with the solution (\ref{ep32}) in normal hierarchy.}
{\begin{tabular}{@{}cccc@{}} \toprule
Parameter & Best fit & $1\sigma $ range& $2\sigma $ range \\
\noalign{\smallskip}\hline\noalign{\smallskip}
$\mathrm{A [eV]}$                  & $10^{-2}$ & $J$ & $-0.00303931$ \\
$\mathrm{B_1[eV]}$                 & $-0.0356733$ & $|m_{1}| [eV]$ & $0.00487852$ \\
$\mathrm{B_2[eV]}$                 & $-0.0199378$ & $|m_{2}|$ & $0.01$   \\
$\mathrm{C [eV]}$                  & $0.0215348$ & $|m_{3}| [eV]$ & $0.0507326$ \\
 $\mathrm{\theta _{13} [^{\circ}]}$& $8.05436$ & $\sum [eV]$ & $0.0656112$ \\
 $\mathrm{\theta_{12}[^{\circ}]}$  & $34.929$ & $|m_{ee}| [eV]$ & $0.0010064$  \\
 $\mathrm{\theta_{23}[^{\circ}]}$  & $44.9281$ & $|m_{\beta}| [eV]$ & $0.00991761$  \\
 $\mathrm{\delta [^{\circ}]}    $  & $354.59$ &  \\\botrule
\end{tabular} \label{resultaN}}
\end{table}

Now, substituting $K=-1.43$ in Eq.(\ref{K}) we obtain \be B_1=B_2-0.730699C. \label{B1B2v1} \ee

\subsection{\label{Normal} Normal hierarchy ($\Delta m^2_{31}> 0$)}
Combining (\ref{B1B2v1}) and (\ref{m123}) with the two
experimental constraints on squared mass differences of neutrinos in normal hierarchy
as shown in Tab.\ref{NormalH}, we get the solutions (in
[eV]) given in \ref{K1N}.
The solutions from Eq. (\ref{case1}) to Eq.(\ref{case4})
have the same absolute values of $m_{1,2, 3}$, the unique
difference is the sign of $m_{1,3}$. Hence, we only consider in
detail the case
of (\ref{case1}). On the other hand, the expressions
from (\ref{case1}) to (\ref{case4}) show that $m_{i}\,\, (i=1,2,3)$
depends only on one parameter $A=m_2$,  so we will consider $m_{1,3}$ as
functions of $m_2$. However, to have an explicit hierarchy on
neutrino masses, in the following figures, $m_2$ should be
included.
The use of the upper bound on absolute value of
neutrino mass in (\ref{upb}) leads to $A \leq 0.6\,\mathrm{eV}$.
Moreover,  in this case,
  $A \in (0.00873, 0.01)\, \mathrm{eV}$ or $A \in (-0.01, -0.00873)\, \mathrm{eV}$ are good
   regions of $A$ that can reach the realistic neutrino mass hierarchy.

\begin{figure}[h]
\bc
\includegraphics{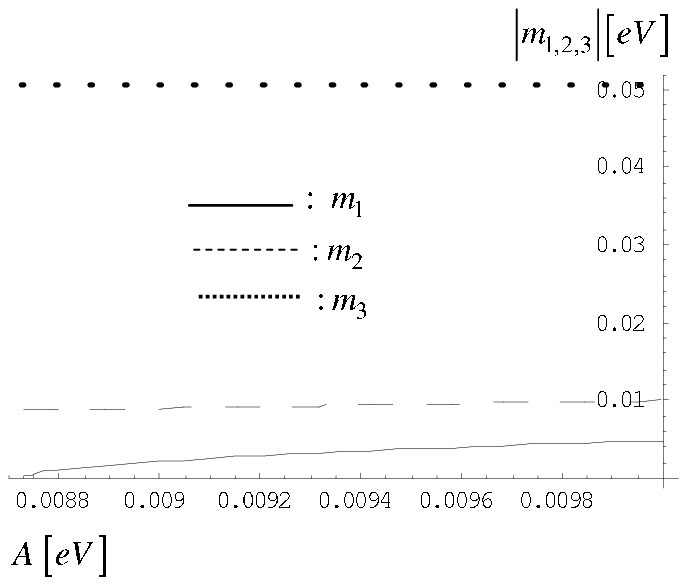}
\vspace*{-0.1cm} \caption[$|m_{1,2,3}|$ as functions of $A$ in
 the case of $\Delta m^2_{31}> 0$ with
 $A\in(0.00873, 0.01) \,\mathrm{eV}$.]{$|m_{1,2,3}|$ as functions of $A$ in
 the case of $\Delta m^2_{31}> 0$ with
 $A\in(0.00873, 0.01) \,\mathrm{eV}$.}\label{m123Ncase1}
\ec
\end{figure}

\begin{figure}[h]
\begin{center}
\includegraphics{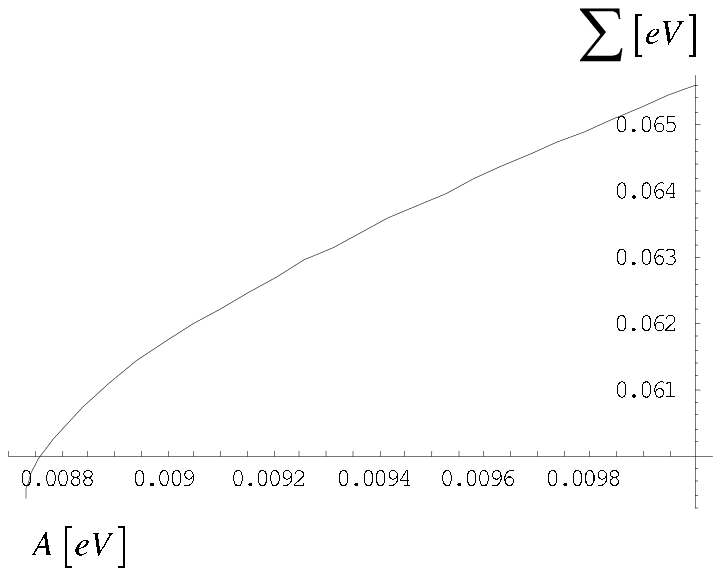}
\vspace*{-0.1cm} \caption[$\sum$ as a function of
$A$ with $A \in (0.00873, 0.01)\,\mathrm{eV}$ in the
case of $\De m^2_{31}> 0$]{$\sum$ as a function of
$A$ with $A \in (0.00873, 0.01)\,\mathrm{eV}$ in the
\textbf{case of $\De m^2_{31}> 0$}.}\label{m123Nscase1}
\end{center}
\end{figure}

In Fig. \ref{m123Ncase1}, we have plotted the absolute values $|m_{1,2,3}|$
as functions of $A$ with
$A \in (0.00873, 0.01)\, \mathrm{eV}$.
 This figure shows
  that there exist allowed regions for values $A$ (or $m_2$) where either normal
  or quasi-degenerate neutrino masses spectrum is achieved.
  The quasi-degenerate mass hierarchy is obtained when $|A|$ lies
   in a region [$0.05\,\mathrm{eV} , +\infty$] ($|A|$ increases
  but must be small enough because of the scale of $m_{1,2,3}$). The normal
  mass hierarchy will be obtained if $|A|$ takes the values around
   $(0.00873, 0.05)\,
   \mathrm{eV}$. The sum
   $\sum=\sum_{i=1}^3|m_i|$ is plotted in Fig. \ref{m123Nscase1}
    with $m_2 \in (0.00873, 0.01)\,\mathrm{eV}$.
\\
The effective mass $\langle m_{ee}\rangle$ governing neutrinoless double beta
 decay \cite{betdecay1,betdecay2,betdecay3,betdecay4,betdecay5} is then obtained,
\bea \langle m_{ee}\rangle &=& \sum^3_{i=1} U_{ei}^2 m_i =(0.321369 + 0.00369042i)A\crn
&-&(0.339313 - 0.00184244i)\sqrt{4A^2-0.0003048}\crn
&-&(0.0205292 - 0.0039231i)\sqrt{\al_1-2\sqrt{\beta_1}},\label{mee} \eea
and
\bea m^2_\beta &=&\sum^3_{i=1} |U_{ei}|^2 m_i^2 =-4.66947\times 10^{-6} 
+1.03963A^2-0.0445038\sqrt{\beta_1} \crn
&+&0.0209007\sqrt{4A^2-0.0003048}\sqrt{\al_1-2\sqrt{\beta_1}},\label{mb}\eea
with $\al_1, \beta_1$ are given in (\ref{albet1}).

\begin{figure}[ht]
\begin{center}
\includegraphics{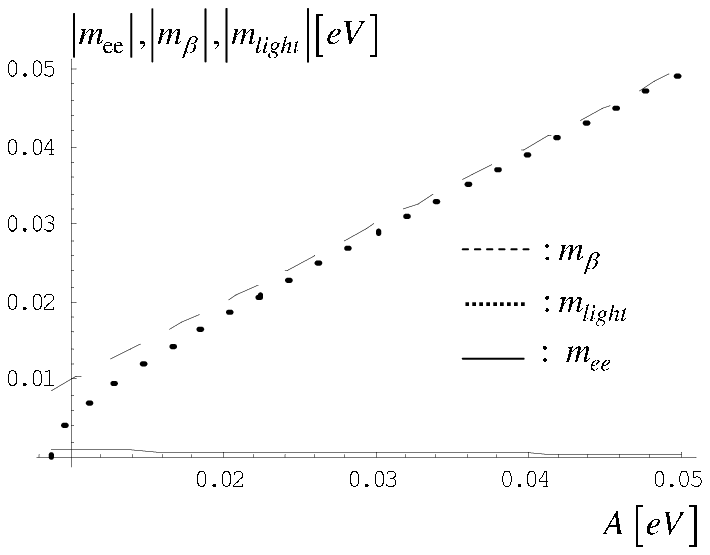}
\vspace*{-0.1cm} \caption[$|m_{ee}|$, $|m_{\beta}|$ and $|m_{light}|$ as functions of $A$
from (\ref{case1}) in normal hierarchy with $A\in(0.00875, 0.05) \,
\mathrm{eV}$]{$|m_{ee}|$, $|m_{\beta}|$ and $|m_{light}|$ as functions of $A$
from (\ref{case1}) in normal hierarchy with $A\in(0.00875, 0.05) \,
\mathrm{eV}$}\label{meeA4}
\end{center}
\end{figure}

We also note that in the normal spectrum, $|m_1|\approx
|m_2|<|m_3|$, so $m_1$ given in (\ref{case1}) is the lightest
neutrino mass, {\rm therefore,}
 it is denoted as $m_{1}\equiv m_{light}$. In
Fig. \ref{meeA4} we have plotted the values of $|m_{ee}|$,
$|m_{\beta}|$ and $|m_{light}|$ as functions of $A$ with
$A\in (0.00875, 0.05)\,\mathrm{eV}$.

Fig. \ref{meeA4} shows that in normal case $\langle m_{ee}\rangle < |m_\beta|< |m_{light}|$,
and all of them are consistent with the recent experimental data\cite{PDG2012} . 
By assuming $A\equiv m_2=10^{-2}\, \mathrm{eV}$, which is safely small,
then the other neutrino masses are explicitly
 given as $m_1=-4.87852\times 10^{-3}\, \mathrm{eV},\, m_3= -5.07326 \times 10^{-2} \,
 \mathrm{eV}$, and $\sum=6.56112\times 10^{-2}\, \mathrm{eV}$, $|m_{ee}|\simeq
 0.900184 \times 10^{-3} \, \mathrm{eV},\, |m_{\beta}|\simeq 0.991761 \times 10^{-2} \,
 \mathrm{eV}$. Three physical neutrino masses are: $|m_1|=4.87852\times 10^{-3}\,
 \mathrm{eV}, \,\,|m_2|=10^{-2} \, \mathrm{eV}$, $|m_3|= 5.07326 \times 10^{-2}$.
  This solution means a normal
neutrino mass spectrum as mentioned above and consistent with the
recent experimental data \cite{PDG2012,Nuexp1} . It follows that
\bea
 B_1&=&-0.0356733 \,\mathrm{eV}, \,\,
 B_2=-0.0199378\,\mathrm{eV},\,\,
 C=0.0215348\,\mathrm{eV}. \label{Vcase1}\eea
 There has not yet been an explicit experimental test of the values of parameters
  $\la_{s,\si}, v_{s,\si}, \La_{s,\si}$, however, from the original form of the 3-3-1 models they
   obey the relation \cite{DL2008} $\la_{s,\si}\sim v^2_{s,\si}/\La_{s,\si}$. To show that there exist
    the model parameters that consist with experimental data, the following assumption is used:
\bea
\la_s&=&\la_{\si}=1\,\mathrm{eV},\,\, v_\rho=v_s=v_\si,\,\,
\La_s=-\La_\si=v^2_\si,\label{assum}\eea
It is then
\bea
 A&=&2x, \,\, B_1=\frac{x(2 x^2 + 2 y^2 - 4 y z + z^2)}{x^2-y^2},\crn
 B_2&=&\frac{x(2 x^2 + 2y^2 + 4 y z + z^2)}{x^2-y^2},\,\, C=\frac{y(4x^2- z^2)}{x^2-y^2}. \hs \label{ABCcasse1}
\eea
Combining (\ref{Vcase1}) and (\ref{ABCcasse1}) yields:
$x=5\times 10^{-3}$, $y\simeq $ $ -7.73 \times10^{-3}$,
\, $z\simeq 1.77\times10^{-3}$.

\subsection{Inverted case ($\Delta m^2_{31}< 0$)}
For inverted hierarchy, by combining (\ref{B1B2v1}) and (\ref{m123}) with the two
experimental constraints on squared mass differences of neutrinos
as shown in Tab.\ref{InvertedH}, we get the solutions (in
[eV]) given in \ref{K1I}.
The solutions from Eq. (\ref{case114I}) to Eq.(\ref{case214I})
have the same absolute values of $m_{1,2, 3}$, the unique
difference is the sign of $m_{1,3}$. Hence, we only consider in
detail the case
of (\ref{case114I}). Because $m_{i}\,\, (i=1,2,3)$ only depends on one parameter $A=m_2$, so we will consider $m_{1,3}$ as
functions of $A$. However, to have an explicit hierarchy on
neutrino masses, in the following figures, $m_2$ should be
included.
In this case, $A \in (0.05, 0.06)\, \mathrm{eV}$ is a
   good region of $A$ that can reach the realistic neutrino mass hierarchy.

\begin{figure}[h]
\bc
\includegraphics{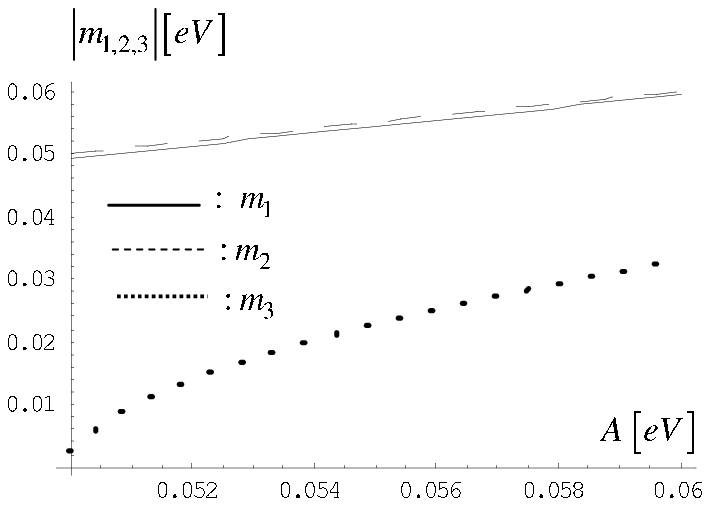}
\vspace*{-0.1cm} \caption[$|m_{1,2,3}|$ as functions of $A$ with
 $A\in(0.05, 0.06) \,\mathrm{eV}$ in
 the case of $\Delta m^2_{31}< 0$]{$|m_{1,2,3}|$ as functions of $A$ with
 $A\in(0.05, 0.06) \,\mathrm{eV}$ in
 the case of $\Delta m^2_{31}< 0$.}\label{m123Icase1}
\ec
\end{figure}

\begin{figure}[h]
\begin{center}
\includegraphics{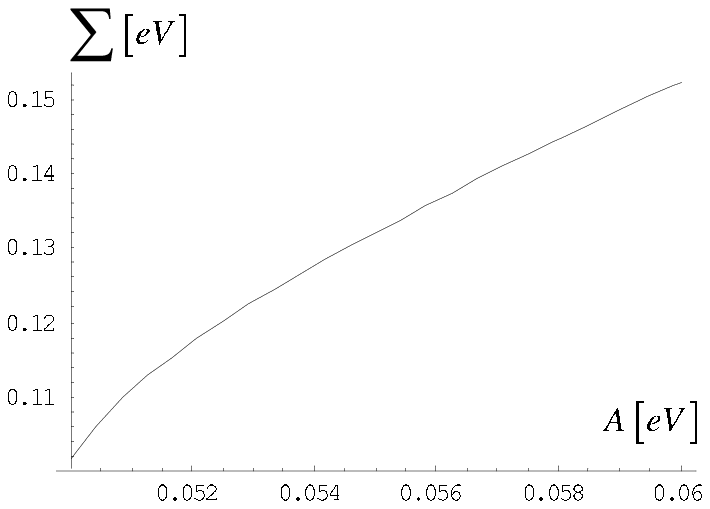}
\vspace*{-0.1cm} \caption[$\sum$ as a function of
$A$ with $A \in (0.05, 0.06)\,\mathrm{eV}$ in the
case of $\De m^2_{31}< 0$]{$\sum$ as a function of
$A$ with $A \in (0.05, 0.06)\,\mathrm{eV}$ in the
case of $\De m^2_{31}< 0$.}\label{m123Iscase1}
\end{center}
\end{figure}

In Fig. \ref{m123Icase1}, we have plotted the absolute values $|m_{1,2,3}|$
as functions of $A$ with
$A \in (0.05, 0.06)\, \mathrm{eV}$.
 This figure shows
  that there exist allowed regions for values $A$ (or $m_2$) where either inverted
  or quasi-degenerate neutrino masses spectrum is achieved.
  The quasi-degenerate mass hierarchy is obtained when $|A|$ lies
   in a region [$0.06\,\mathrm{eV} , +\infty$] ($|A|$ increases
  but must be small enough because of the scale of $m_{1,2,3}$). The inverted
  mass hierarchy will be obtained if $|A|$ takes the values around
   $(0.05, 0.06)\,
   \mathrm{eV}$. The sum
   $\sum=\sum_{i=1}^3|m_i|$ is plotted in Fig. \ref{m123Iscase1}
    with $m_2 \in (0.05, 0.06)\,\mathrm{eV}$.

The effective mass $\langle m_{ee}\rangle$ governing neutrinoless
double beta decay \cite{betdecay1,betdecay2,betdecay3,betdecay4,betdecay5} in inverted hierarchy
is then obtained,
\bea \langle m^I_{ee}\rangle &=& \left|\sum^3_{i=1} U_{ei}^2 m_i \right|=\left|(0.321369 +
 0.00369042i)A\right.\crn
&+&\left.(0.339313- 0.00184244i)\sqrt{4A^2-0.0003048}\right.\crn
&-&\left.(0.0205292 - 0.0039231i)\sqrt{\al_3-2\sqrt{\beta_3}}\right|, \label{meeI}\eea
and
\bea (m^{I}_\beta)^2 &=&\sum^3_{i=1} |U_{ei}|^2 m_i^2 =-0.000102434+1.03963A^2 
+0.0445038\sqrt{\beta} \crn
&-&0.0209007\sqrt{4A^2-0.0003048}\sqrt{\al_3-2\sqrt{\beta_3}},\label{mbI}\eea
with $\al_3, \beta_3$ are given in (\ref{albet3}).

We also note that in the inverted spectrum, $|m_2|\approx
|m_1|>|m_3|$, so $m_3$ given in (\ref{case114I}) is the lightest
neutrino mass, {\rm therefore,}
it is denoted as $m_{3}\equiv m^I_{light}$. In
Fig. \ref{meeA4} we have plotted the values of $|m^I_{ee}|$,
$|m^I_{\beta}|$ and $|m^I_{light}|$ as functions of $A$ with
$A\in (0.05, 0.06)\,\mathrm{eV}$.
 \begin{figure}[ht]
\begin{center}
\includegraphics{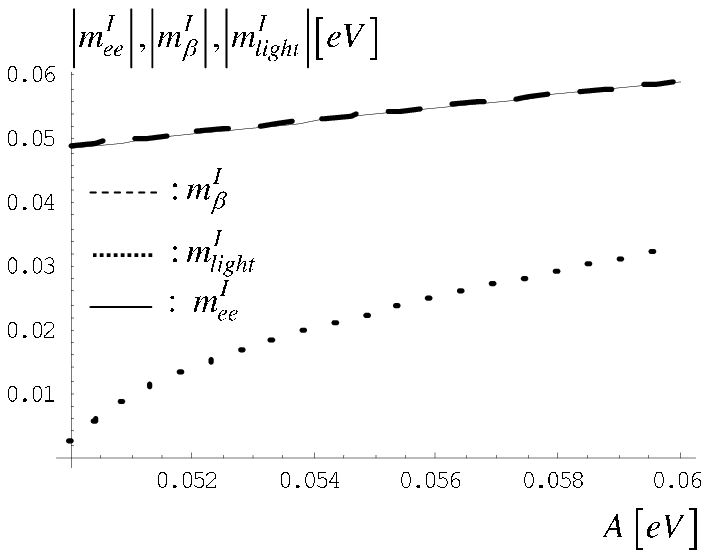}
\vspace*{-0.1cm} \caption[$|m^I_{ee}|$, $|m^I_{\beta}|$ and $|m^I_{light}|$ as functions of $A$
from (\ref{case114I}) in inverted hierarchy with $A\in(0.05, 0.06) \,
\mathrm{eV}$]{$|m^I_{ee}|$, $|m^I_{\beta}|$ and $|m^I_{light}|$ as functions of $A$
from (\ref{case114I}) in inverted hierarchy with $A\in(0.05, 0.06) \,
\mathrm{eV}$}\label{meeIA4}
\end{center}
\end{figure}

By assuming $A\equiv m_2=5\times 10^{-2}\, \mathrm{eV}$, which is safely small,
then the other physical neutrino masses are explicitly
 given as $|m_1|=4.92321\times 10^{-2}\, \mathrm{eV},\, |m_3|= 2.48998\times 10^{-3} \,
 \mathrm{eV}$, and $\sum=0.101722 \, \mathrm{eV}$, $m^I_{ee}\simeq
 4.85391\times 10^{-2} \, \mathrm{eV}$, $m^I_{\beta}\simeq 4.90048 \times 10^{-2} \, \mathrm{eV}$.
  This solution means a inverted neutrino mass spectrum as mentioned above and consistent with the
recent experimental data \cite{PDG2012,Nuexp1} . It follows that
\bea
 B_1&=&(1.61687+ 0.167223i)\times 10^{-2} \,\mathrm{eV}, \crn
 B_2&=&(3.30634+ 0.0817754 i)\times 10^{-2}\mathrm{eV},\crn
 C&=&(2.31213 - 0.116939i)\times 10^{-2}.\,\mathrm{eV} \label{Vcase1I}\eea
Furthermore, by assuming that \cite{DL2008}
\bea
\la_s&=&\la_{\si}=1\,\mathrm{eV},\,\, v_\rho=v_s=v_\si,\,\,
\La_s=a \La_\si,\,\, \La_\si=-v^2_\si,\label{assumI}\eea
we obtain a solution
\bea
 A&=&2x, \,\, C=\frac{(a+3)x}{a+1}+\frac{a z^2}{x(1-a^2)},\, x=y, \crn
 B_1&=&\frac{(a-1)(a+3)x^2 - 2(a-1) x z - z^2}{x(a^2-1)},\crn
 B_2&=&\frac{(a-1)(a+3)x^2 + 2(a-1) x z - z^2}{x(a^2-1)}. \hs \label{ABCcasse1I}
\eea
Combining (\ref{Vcase1I}) and (\ref{ABCcasse1I}) yields:
\bea
a\simeq 1.0587 - 0.0097i, \,\,
x=y=2.5\times 10^{-2},\, z\simeq (8.6933- 0.4808i)\times10^{-3}.\eea

\section{\label{quarkA4} Quarks sector}
We note that the scalar triplet $\phi$ in Eq. (\ref{phi}) is not enough to generate mass for all the quarks. Hence, to generate masses for quarks,  two $SU(3)_L$ triplets, put in $\underline{1}$ and $\underline{3}$ under $A_4$, are additional introduced \cite{dlshA4} :
\bea \eta&=&
\left(\eta^0_1 \hs
  \eta^-_2 \hs
  \eta^0_3\right)^T\sim (3,-1/3,-1/3, \underline{3}),\label{eta}\\
  \chi&=&\left(\chi^0_1 \hs
  \chi^-_2 \hs
  \chi^0_3 \right)^T\sim (3,-1/3,2/3,\underline{1}).\label{chi}\eea 
It is worth mentioning that the $SU(3)_L$ triplet
$\rho$ does not give new Yukawa terms, so the results in quark sector remain the same. The Yukawa
interactions are: \bea -\mathcal{L}_q &=& h^d_3 \overline{Q}_{3L}(\phi
d_R)_1
 + h^u_1 \overline{Q}_{1L}(\phi^*u_R)_{1''}
 + h^u_2
\overline{Q}_{2L}(\phi^*u_R)_{1'}\crn
&+& h^u_3 \overline{Q}_{3L}(\eta
u_R)_1
+h^d_1 \overline{Q}_{1L}(\eta^* d_R)_{1''}
+h^d_2
\overline{Q}_{2L}(\eta^* d_R)_{1'}\crn
&+& f_3 \overline{Q}_{3L}\chi U_R +
f_1 \overline{Q}_{1L}\chi^* D_{1R}
+f_2 \overline{Q}_{2L}\chi^* D_{2R}
+H.c.\label{Lquark}\eea
The VEVs of $\eta$ and $\chi$ are supposed to be \cite{dlshA4}
\bea
\langle \eta\rangle= (\langle \eta_1\rangle,\langle \eta_1\rangle,\langle \eta_1\rangle) \label{eta}\eea
 under $A_4$, where
 $\langle \eta_1\rangle=(u\,\,\,\, 0\,\,\,\, 0)^T$ and $\langle \chi\rangle=(0\,\,\,\, 0\,\,\,\, v_\chi)^T$.
The exotic quarks get masses directly
from the VEV of $\chi$ \cite{dlshA4}: $m_U=f_3 v_\chi$, $m_{D_{i}}=f_{i}
v_\chi, \,\, (i=1,2)$. 

Substituting (\ref{phi}),(\ref{epsi12}) and (\ref{eta}) into (\ref{Lquark}), the mass matrices
for ordinary up-quarks and
down-quarks are, respectively, obtained as follows:
\bea M_u &=&
\left(%
\begin{array}{ccc}
  -h^u_1 v & -\om h^u_1  v (1+\varep_2)  & -\om^2 h^u_1 v (1+\varep_2) \\
   -h^u_2 v & -\om^2 h^u_2 v(1+\varep_2)  & -\om h^u_2 v(1+\varep_2)   \\
  h^u_3 u & h^u_3 u & h^u_3 u \\
\end{array}%
\right),\hs \label{Mu}\\
M_d&=&
\left(%
\begin{array}{ccc}
  h^d_1 u & \om h^d_1 u  & \om^2 h^d_1 u  \\
   h^d_2u & \om^2 h^d_2 u  & \om h^d_2 u   \\
  h^d_3 v & h^d_3 v(1+\varep_2) & h^d_3 v(1+\varep_3) \\
\end{array}%
\right).\label{Md} \eea
The matrices $M_u$ and $M_d$ in (\ref{Mu}), (\ref{Md}) are, respectively, diagonalized as
\bea U^{u+}_L M_u U^u_R&=&\mathrm{diag} (m_u,\,\,m_c,\,\, m_t), \crn
 U^{d+}_L M_d U^d_R&=&\mathrm{diag}(m_d,\,\, m_s,\,\, m_b),\eea
where
\bea
m_u&=&-\frac{(1+i\sqrt{3})(3+2\varep_2+2\varep_3+\varep_2\varep_3)h^u_1 v}{3i+\sqrt{3}+(i+\sqrt{3})+2i\varep_3},\crn
m_c&=&\frac{(-1+i\sqrt{3})(3+2\varep_2+2\varep_3+\varep_2\varep_3)h^u_2 v}{-3i+\sqrt{3}+(-i+\sqrt{3})\varep_3},\crn
m_t&=&\frac{(3+2\varep_2+2\varep_3+\varep_2\varep_3)h^u_3 u}{\sqrt{3}(1+\varep_2)},\,\,
m_d=\frac{(1+i\sqrt{3})(3+2\varep_2+2\varep_3)h^d_1 u}{3i+\sqrt{3}+(i+\sqrt{3})\varep_3+2i\varep_2},\crn
m_s&=&\frac{(1-i\sqrt{3})(3+\varep_2+\varep_3)h^d_2 u}{-3i+\sqrt{3}-2i\varep_3},\,\,
m_b=\frac{(3+\varep_2+\varep_3)h^d_3 v}{\sqrt{3}},\label{mdsb}
\eea
and $U^u_R, U^d_R$ are the right-handed
up- and down -quarks mixing matrices; $U^u_L, U^d_L$ are the left-handed
up- and down-quarks mixing matrices, respectively,
\bea U^u_R&=&
\fr{1}{\sqrt{3}}\left(%
\begin{array}{ccc}
  1 &\frac{\om^2(1+\varep_2)-(1+\varep_3)}{\om^2(1+\varep_3)-1} & 1+\varep_3 \\
  \frac{\om(1+\varep_3)-1}{\om^2(1+\varep_2)-\om(1+\varep_3)} & \om & \frac{1+\varep_3}{1+\varep_2}\\
  \frac{1-\om^2(1+\varep_2)}{\om^2(1+\varep_2)-\om(1+\varep_3)} & \frac{\om(1-\om)-\om^2 \varep_2}{\om^2(1+\varep_3)-1}  & 1 \\
\end{array}%
\right),\crn
U^d_R&=&
\fr{1}{\sqrt{3}}\left(%
\begin{array}{ccc}
  1 &\frac{\om^2(1+\varep_3)-(1+\varep_2)}{\om^2-(1+\varep_3)} & 1\\
  \frac{\om-(1+\varep_3)}{\om^2(1+\varep_3)-\om(1+\varep_2)} & \om & 1\\
  \frac{1+\varep_2-\om^2}{\om^2(1+\varep_3)-\om(1+\varep_2)} & \frac{\om(1-\om)+\om \varep_2}{\om^2-(1+\varep_3)}  & 1 \\
\end{array}%
\right),\label{UduR}\\
U^u_L&=&U^d_L=1.\nn\eea
The right-handed
up- and down -quarks mixing matrices $U^u_R, U^d_R$ given in (\ref{UduR}) is one of the different issues of this work compared with Ref.\citen{dlshA4} . However, the CKM matrix is then given as \cite{PDG2012} \bea
U_\mathrm{CKM}= U^u_LU^{d\dagger}_L=1,\label{a41}\eea
which is the same as that in Ref. \citen{dlshA4} . A tree-level CKM matrix obtained equal to the identity matrix is the common property for some models based on the $A_4$ group \cite{dlsvS4} .

\section{\label{conclus}Conclusions}

In this paper, we have proposed a 3-3-1 model with neutral fermions based on $A_4$ flavor symmetry
responsible for fermion masses and mixings with non-zero $\theta_{13}$. For this purpose, we
additionally introduce  a new  $SU(3)_L$ triplet ($\rho$) lying in $\underline{3}$ under $A_4$. The neutrinos get
small masses from two $SU(3)_L$ anti-sextets and one $SU(3)_L$ triplet. The model can fit the present data on neutrino masses and mixing as well as the effective mass governing neutrinoless
 double beta decay.
 Our results show that the neutrino masses are naturally small and a little deviation from the
tri-bimaximal neutrino mixing form can be realized. The Dirac CP violation phase $\delta$ is predicted to either
  $5.41^{\circ}$ or $354.59^{\circ}$ with $\theta_{23} \neq  \fr \pi 4$. It is emphasized that this consequence does not require
 $\theta_{23} = \fr \pi 4$.

 \section*{Acknowledgments}
This research is funded by Vietnam  National
Foundation for Science and Technology Development (NAFOSTED) under
grant number 103.01-2014.51.

\appendix

\section{\label{apa}$\emph{A}_4$ group and Clebsch-Gordan coefficients }

$A_4$ is the group of even
permutation of four objects, which is also the symmetry group of a
regular tetrahedron. It has 12 elements and four equivalence classes with three
inequivalent one-dimensional representations and one
three-dimensional one. Any element of $A_4$ can be formed by
multiplication of the generators $S$ and $T$ obeying the relations\cite{dlshA4}
$S^{2}=T^{3}=(ST)^{3}={\bf 1}$ . Without loss of generality, we could
choose $S = (12)(34),\ T = (234)$ where the cycles (12)(34) denotes the
permutation $(1, 2, 3, 4) \rightarrow (2, 1, 4, 3)$, and (234)
means $(1, 2, 3, 4) \rightarrow (1, 3, 4, 2)$. The conjugacy
classes of $A_4$ generated from $S$ and $T$ are
\bea C_1 &:& 1 \crn
 C_2 &:& S, \,TST^2,\, T^2 ST, \crn
 C_3 &:&T, \, TS, \, ST, \, STS,\crn
  C_4 &:&T^2,\, ST^2, \, T^2S,\, TST.\nn \eea

The character table of $A_4$ is given in Table \ref{A4table}, where $n$ is the
 order of class and $h$ the order of elements
within each class.
\begin{table}[h]
\caption{\label{A4table}The character table of $A_4$ group}
\begin{center}
\begin{tabular}{lllllll}
\hline\noalign{\smallskip}
 Class & $n$ & $h$&$\chi_{\underline{1}}$&$\chi_{\underline{1}'}$
 &$\chi_{\underline{1}''}$&$\chi_{\underline{3}}$\\
\noalign{\smallskip}\hline\noalign{\smallskip}
$C_{1}$ &1 & 1& 1& 1& 1& 3 \\
$C_{2}$ &3 & 2& 1& 1& 1& -1 \\
$C_{3}$ &4 & 3& 1& $\om$& $\om^2$& 0 \\
$C_{4}$ &4 & 3& 1& $\om^2$& $\om$& 0 \\
\noalign{\smallskip}\hline
\end{tabular}
\end{center}
\end{table}
We will work on a basis where $\underline{3}$ is a real representation. One
possible choice of generators is given as follows \bea
\underline{1}&:& S=1,\hs T=1 \crn
\underline{1}'&:& S=1,\hs T=\om \crn
\underline{1}'' &:& S=1,\hs T=\om^2,\crn
\underline{3}&:& S=\left(%
\begin{array}{ccc}
  1 & 0 & 0 \\
  0 & -1 & 0 \\
  0 & 0 & -1 \\
\end{array}%
\right),\hs T=\left(%
\begin{array}{ccc}
  0 & 1 & 0 \\
  0 & 0 & 1 \\
 1 & 0 & 0 \\
\end{array}%
\right),\nn\eea where $\om=e^{2\pi i/3}=-1/2+i\sqrt{3}/2$ is the
cube root of unity. Using them we calculate the Clebsch-Gordan
coefficients for all the tensor products as given below.

First, let us put $\underline{3}(1,2,3)$ which means some
$\underline{3}$ multiplet such as $x=(x_1,x_2,x_3)\sim
\underline{3}$ or $y=(y_1,y_2,y_3)\sim \underline{3}$ and so on,
and similarly for the other representations. Moreover, the
numbered multiplets such as $(...,ij,...)$ mean $(...,x_i
y_j,...)$ where $x_i$ and $y_j$ are the multiplet components of
different representations $x$ and $y$, respectively. In the
following the components of representations in l.h.s will be
omitted and should be understood, but they always exist in order
in the components of decompositions in r.h.s: \bea
\underline{1}\otimes\underline{1}&=&\underline{1}(11),\,
\underline{1}\otimes \underline{1}'=\underline{1}'(11),\,
\underline{1}\otimes\underline{1}''=\underline{1}''(11),\\
\underline{1}'\otimes \underline{1}'&=&\underline{1}''(11),\,
\underline{1}'\otimes \underline{1}''=\underline{1}(11),\,
\underline{1}''\otimes \underline{1}''=\underline{1}'(11),\\
\underline{1}\otimes \underline{3}&=&\underline{3}(11,12,13),\,
\underline{1}'\otimes \underline{3}=\underline{3}(11,\om12,\om^213),\crn
\underline{1}''\otimes \underline{3}&=&\underline{3}(11,\om^212,\om13),\\
\underline{3} \otimes \underline{3} &=& \underline{1}(11+22+33)
\oplus \underline{1}'(11+\om^2 22+ \om 33)\oplus
\underline{1}''(11+\om 22+ \om^2 33) \crn &\oplus& \underline{3}_s
(23+32,31+13,12+21)\oplus
\underline{3}_a(23-32,31-13,12-21),\eea where the subscripts
$"s"$ and $"a"$ respectively refer to their symmetric and
antisymmetric product combinations as explicitly pointed out.

In the text we usually use the following notations, for example,
$(xy')_{\underline{3}}=
[xy']_{\underline{3}}\equiv(x_2y'_3-x_3y'_2,x_3y'_1-x_1y'_3,x_1y'_2-x_2y'_1)$
which is the Clebsch-Gordan coefficients of $\underline{3}_a$ in
the decomposition of $\underline{3}\otimes \underline{3}$, where
as mentioned $x=(x_1,x_2,x_3)\sim \underline{3}$ and
$y'=(y'_1,y'_2,y'_3)\sim \underline{3}$.

The rules to conjugate the representations \underline{1},
\underline{1}$'$, \underline{1}$''$ and
\underline{3} are given by \bea
&&\underline{1}^*(1^*)=\underline{1}(1^*),\,
\underline{1}'^*(1^*)=\underline{1}'(1^*),\, \underline{1}''^*(1^*)=\underline{1}''(1^*),\\
&&\underline{3}^*(1^*,2^*,3^*)=\underline{3}(1^*,2^*,3^*).\nn\eea

\section{\label{apb} Lepton Number and Lepton Parity}
The lepton number ($L$) and lepton parity ($P_l$) of the model particles are given
 in Tab. \ref{numbers}.
\begin{table}[h]
\tbl{The model particles.}
{\begin{tabular}{@{}cccc@{}} \toprule
$\mathrm{Particles}$ & $L$ & $P_l$\\
\noalign{\smallskip}\hline\noalign{\smallskip}
$N_R$, $u$, $d$,  $\phi^+_1$,$\phi'^+_1$, $\phi^0_2$,$\phi'^0_2$,
$\eta^0_1$,$\eta'^0_1$, $\eta^-_2$,$\eta'^-_2$,$\chi^0_3$, $\sigma^0_{33}$,
$s^0_{33}$ &0 &1\\\noalign{\smallskip}\hline
$\nu_L$, $l$, $U$, $D^*$, $\phi^+_3$,$\phi'^+_3$, $\eta^0_3$,$\eta'^0_3$, $\chi^{0*}_1$,
$\chi^+_2$,$\sigma^0_{13}$, $\sigma^+_{23}$, $s^0_{13}$, $s^+_{23}$&-1
&-1\\\noalign{\smallskip}\hline
$\sigma^{0}_{11}$, $\sigma^{+}_{12}$, $\sigma^{++}_{22}$,  $s^{0}_{11}$,
$s^{+}_{12}$, $s^{++}_{22}$ &-2&1 \\ \botrule
\end{tabular} \label{numbers}}
\end{table}

\section{\label{K1N}The solutions with $K=-1.43$ in the normal case}
 \begin{itemize}
 \item The first case:
 \bea
C&=&0.5\sqrt{\al_1-2\sqrt{\beta_1}},\crn
B_2&=&-0.5\sqrt{4 A^2-0.0003048}
-0.34965\sqrt{\al_1-2\sqrt{\beta_1}},\crn
m_1&=&-0.5\sqrt{4 A^2-0.0003048}\crn
&-&1.11022\times 10^{-16}\sqrt{\al_1-2\sqrt{\beta_1}},\,\,\,\, m_2=A, \label{case1}\\
m_3&=&-0.5\sqrt{4 A^2-0.0003048}-1.06465\sqrt{\al_1-2\sqrt{\beta_1}}.\nn \eea
\item The second case:
  \bea
C&=&0.5\sqrt{\al_1+2\sqrt{\beta_1}},\crn
B_2&=&-0.5\sqrt{4 A^2-0.0003048}
-0.34965\sqrt{\al_1+2\sqrt{\beta_1}},\crn
m_1&=&-0.5\sqrt{4 A^2-0.0003048}\crn
&-&1.11022\times 10^{-16}\sqrt{\al_1+2\sqrt{\beta_1}},\,\,\,\, m_2=A, \label{case2}\\
m_3&=&-0.5\sqrt{4 A^2-0.0003048}-1.06465\sqrt{\al_1+2\sqrt{\beta_1}}.\nn \eea
\item The third case:
 \bea
C&=&0.5\sqrt{\al_1-2\sqrt{\beta_1}},\crn
B_2&=&0.5\sqrt{4 A^2-0.0003048}
-0.34965\sqrt{\al_1-2\sqrt{\beta_1}},\crn
m_1&=&0.5\sqrt{4 A^2-0.0003048}\crn
&-&1.11022\times 10^{-16}\sqrt{\al_1-2\sqrt{\beta_1}},\,\,\,\, m_2=A, \label{case3}\\
m_3&=&0.5\sqrt{4 A^2-0.0003048}-1.06465\sqrt{\al_1-2\sqrt{\beta_1}}.\nn
\eea
\item The fourth case:
 \bea
C&=&0.5\sqrt{\al_1+2\sqrt{\beta_1}},\crn
B_2&=&0.5\sqrt{4 A^2-0.0003048}
-0.34965\sqrt{\al_1+2\sqrt{\beta_1}},\crn
m_1&=&0.5\sqrt{4 A^2-0.0003048}\crn
&-&1.11022\times 10^{-16}\sqrt{\al_1+2\sqrt{\beta_1}},\,\,\,\, m_2=A, \label{case4}\\
m_3&=&0.5\sqrt{4 A^2-0.0003048}-1.06465\sqrt{\al_1+2\sqrt{\beta_1}}.\nn
\eea
where
\bea
\al_1 &=&0.00211525 + 1.76448 A^2,\label{albet1}\\
\beta_1&=&-1.46721\times 10^{-7} + 0.00186616 A^2 +
    0.778345A^4.\nn \eea
\end{itemize}
\section{\label{K2N}The solutions with $K=1.43$ in the normal case}
 \begin{itemize}
 \item The first case:
 \bea
C&=&-0.5\sqrt{\al_1-2\sqrt{\beta_1}},\crn
B_2&=&-0.5\sqrt{4 A^2-0.0003048}
-0.34965\sqrt{\al_1-2\sqrt{\beta_1}},\crn
m_1&=&-0.5\sqrt{4 A^2-0.0003048}\crn
&-&1.11022\times 10^{-16}\sqrt{\al_1-2\sqrt{\beta_1}},\,\,\,\, m_2=A, \label{case114p}\\
m_3&=&-0.5\sqrt{4 A^2-0.0003048}-1.06465\sqrt{\al_1-2\sqrt{\beta_1}}.\nn \eea
\item The second case:
  \bea
C&=&-0.5\sqrt{\al_1+2\sqrt{\beta_1}},\crn
B_2&=&-0.5\sqrt{4 A^2-0.0003048}
-0.34965\sqrt{\al_1+2\sqrt{\beta_1}},\crn
m_1&=&-0.5\sqrt{4 A^2-0.0003048}\crn
&-&1.11022\times 10^{-16}\sqrt{\al_1+2\sqrt{\beta_1}},\,\,\,\, m_2=A, \label{case214p}\\
m_3&=&-0.5\sqrt{4 A^2-0.0003048}-1.06465\sqrt{\al_1+2\sqrt{\beta_1}}.\nn \eea
\item The third case:
 \bea
C&=&-0.5\sqrt{\al_1-2\sqrt{\beta_1}},\crn
B_2&=&0.5\sqrt{4 A^2-0.0003048}
-0.34965\sqrt{\al_1-2\sqrt{\beta_1}},\crn
m_1&=&0.5\sqrt{4 A^2-0.0003048}\crn
&-&1.11022\times 10^{-16}\sqrt{\al_1-2\sqrt{\beta_1}},\,\,\,\, m_2=A, \label{case314p}\\
m_3&=&0.5\sqrt{4 A^2-0.0003048}-1.06465\sqrt{\al_1-2\sqrt{\beta_1}}.\nn
\eea
\item The fourth case:
 \bea
C&=&-0.5\sqrt{\al_1+2\sqrt{\beta_1}},\crn
B_2&=&0.5\sqrt{4 A^2-0.0003048}
-0.34965\sqrt{\al_1+2\sqrt{\beta_1}},\crn
m_1&=&0.5\sqrt{4 A^2-0.0003048}\crn
&-&1.11022\times 10^{-16}\sqrt{\al_1+2\sqrt{\beta_1}},\,\,\,\, m_2=A, \label{case414p}\\
m_3&=&0.5\sqrt{4 A^2-0.0003048}-1.06465\sqrt{\al_1+2\sqrt{\beta_1}},\nn
\eea
where $\al_1, \beta_1$ are given in (\ref{albet1}).
\end{itemize}
\section{\label{K3N}The solutions with $K=-0.70$ in the normal case}
 \begin{itemize}
 \item The first case:
 \bea
C&=&0.5\sqrt{\al_2-2\sqrt{\beta_2}},\crn
B_2&=&-0.5\sqrt{4 A^2-0.0003048}
-0.714286\sqrt{\al_2-2\sqrt{\beta_2}},\crn
m_1&=&-0.5\sqrt{4 A^2-0.0003048},\,\,\,\, m_2=A, \label{case107}\\
m_3&=&-0.5\sqrt{4 A^2-0.0003048}-1.06429\sqrt{\al_2-2\sqrt{\beta_2}}.\nn \eea
\item The second case:
  \bea
C&=&0.5\sqrt{\al_2+2\sqrt{\beta_2}},\crn
B_2&=&-0.5\sqrt{4 A^2-0.0003048}
-0.714286\sqrt{\al_2+2\sqrt{\beta_2}},\crn
m_1&=&-0.5\sqrt{4 A^2-0.0003048},\,\,\,\, m_2=A, \label{case207}\\
m_3&=&-0.5\sqrt{4 A^2-0.0003048}-1.06429\sqrt{\al_2+2\sqrt{\beta_2}}.\nn \eea
\item The third case:
 \bea
C&=&0.5\sqrt{\al_2-2\sqrt{\beta_2}},\crn
B_2&=&0.5\sqrt{4 A^2-0.0003048}
-0.714286\sqrt{\al_2-2\sqrt{\beta_2}},\crn
m_1&=&0.5\sqrt{4 A^2-0.0003048},\,\,\,\, m_2=A, \label{case307}\\
m_3&=&0.5\sqrt{4 A^2-0.0003048}-1.06429\sqrt{\al_2-2\sqrt{\beta_2}}.\nn
\eea
\item The fourth case:
  \bea
C&=&0.5\sqrt{\al_2+2\sqrt{\beta_2}},\crn
B_2&=&0.5\sqrt{4 A^2-0.0003048}
-0.714286\sqrt{\al_2+2\sqrt{\beta_2}},\crn
m_1&=&0.5\sqrt{4 A^2-0.0003048},\,\,\,\, m_2=A, \label{case407}\\
m_3&=&0.5\sqrt{4 A^2-0.0003048}-1.06429\sqrt{\al_2+2\sqrt{\beta_2}},\nn
\eea
where
\bea
\al_2&=&0.0021167+ 1.76569A^2,\label{albet2}\\
\beta_2&=&-1.46922\times 10^{-7}+0.00186872A^2+0.779412 A^4.\nn
\eea
\end{itemize}

\section{\label{K4N}The solutions with $K=0.70$ in the normal case}
 \begin{itemize}
 \item The first case:
  \bea
C&=&-0.5\sqrt{\al_2-2\sqrt{\beta_2}},\crn
B_2&=&-0.5\sqrt{4 A^2-0.0003048}
-0.714286\sqrt{\al_2-2\sqrt{\beta_2}},\crn
m_1&=&-0.5\sqrt{4 A^2-0.0003048},\,\,\,\, m_2=A, \label{case107p}\\
m_3&=&-0.5\sqrt{4 A^2-0.0003048}-1.06429\sqrt{\al_2-2\sqrt{\beta_2}}.\nn
\eea
\item The second case:
  \bea
C&=&-0.5\sqrt{\al_2+2\sqrt{\beta_2}},\crn
B_2&=&-0.5\sqrt{4 A^2-0.0003048}
-0.714286\sqrt{\al_2+2\sqrt{\beta_2}},\crn
m_1&=&-0.5\sqrt{4 A^2-0.0003048},\,\,\,\, m_2=A, \label{case207p}\\
m_3&=&-0.5\sqrt{4 A^2-0.0003048}-1.06429\sqrt{\al_2+2\sqrt{\beta_2}}.\nn \eea
\item The third case:
 \bea
C&=&-0.5\sqrt{\al_2-2\sqrt{\beta_2}},\crn
B_2&=&0.5\sqrt{4 A^2-0.0003048}
-0.714286\sqrt{\al_2-2\sqrt{\beta_2}},\crn
m_1&=&0.5\sqrt{4 A^2-0.0003048},\,\,\,\, m_2=A, \label{case307p}\\
m_3&=&0.5\sqrt{4 A^2-0.0003048}-1.06429\sqrt{\al_2-2\sqrt{\beta_2}}.\nn
\eea
\item The fourth case:
 \bea
C&=&-0.5\sqrt{\al_2+2\sqrt{\beta_2}},\crn
B_2&=&0.5\sqrt{4 A^2-0.0003048}
-0.714286\sqrt{\al_2+2\sqrt{\beta_2}},\crn
m_1&=&0.5\sqrt{4 A^2-0.0003048},\,\,\,\, m_2=A, \label{case407p}\\
m_3&=&0.5\sqrt{4 A^2-0.0003048}-1.06429\sqrt{\al_2+2\sqrt{\beta_2}},\nn
\eea
where $\al_2, \beta_2$ are given in (\ref{albet2}).
\end{itemize}
 \section{\label{K1I} The solutions with $K=-1.43$ in the inverted case}
 \bit
 \item The first case:
 \bea
C&=&0.5\sqrt{\al_3-2\sqrt{\beta_3}},\crn
B_2&=&0.5\sqrt{4 A^2-0.0003048}
-0.34965\sqrt{\al_3-2\sqrt{\beta_3}},\crn
m_1&=&0.5\sqrt{4 A^2-0.0003048}\crn
&-&1.11022\times 10^{-16}\sqrt{\al_3-2\sqrt{\beta_3}},\,\,\,\, m_2=A, \label{case114I}\\
m_3&=&0.5\sqrt{4 A^2-0.0003048}-1.06465\sqrt{\al_3-2\sqrt{\beta_3}}.\nn
\eea
\item The second case:
 \bea
C&=&0.5\sqrt{\al_3+2\sqrt{\beta_3}},\crn
B_2&=&0.5\sqrt{4 A^2-0.0003048}
-0.34965\sqrt{\al_3+2\sqrt{\beta_3}},\crn
m_1&=&0.5\sqrt{4 A^2-0.0003048}\crn
&-&1.11022\times 10^{-16}\sqrt{\al_3+2\sqrt{\beta_3}},\,\,\,\, m_2=A, \label{case214I}\\
m_3&=&0.5\sqrt{4 A^2-0.0003048}-1.06465\sqrt{\al_3+2\sqrt{\beta_3}},\nn \eea
where
\bea
\al_3 &=&-0.00227829 + 1.76448A^2,\label{albet3}\\
\beta_3&=&1.48642\times 10^{-7}-0.00201 A^2+0.778345 A^4.\nn \eea
\eit
\section{\label{K2I}The solutions with $K=1.43$ in the inverted case}
 \bit
 \item The first case:
 \bea
C&=&-0.5\sqrt{\al_3-2\sqrt{\beta_3}},\crn
B_2&=&0.5\sqrt{4 A^2-0.0003048}
-0.34965\sqrt{\al_3-2\sqrt{\beta_3}},\crn
m_1&=&0.5\sqrt{4 A^2-0.0003048}\crn
&-&1.11022\times 10^{-16}\sqrt{\al_3-2\sqrt{\beta_3}},\,\,\,\, m_2=A, \label{case114pI}\\
m_3&=&0.5\sqrt{4 A^2-0.0003048}-1.06465\sqrt{\al_3-2\sqrt{\beta_3}}.\nn\eea
\item The second case:
 \bea
C&=&-0.5\sqrt{\al_3+2\sqrt{\beta_3}},\crn
B_2&=&0.5\sqrt{4 A^2-0.0003048}
-0.34965\sqrt{\al_3+2\sqrt{\beta_3}},\crn
m_1&=&0.5\sqrt{4 A^2-0.0003048}\crn
&-&1.11022\times 10^{-16}\sqrt{\al_3+2\sqrt{\beta_3}},\,\,\,\, m_2=A, \label{case214pI}\\
m_3&=&0.5\sqrt{4 A^2-0.0003048}-1.06465\sqrt{\al_3+2\sqrt{\beta_3}},\nn \eea
where $\al_3, \beta_3$ are given in (\ref{albet3}).
\eit
\section{\label{K3I} The solutions with $K=-0.7$ in the inverted case}
 \bit
 \item The first case:
 \bea
C&=&0.5\sqrt{\al_4-2\sqrt{\beta_4}},\crn
B_2&=&0.5\sqrt{4 A^2-0.0003048}
-0.714286\sqrt{\al_4-2\sqrt{\beta_4}},\crn
m_1&=&0.5\sqrt{4 A^2-0.0003048},\,\,\,\, m_2=A, \label{case107I}\\
m_3&=&0.5\sqrt{4 A^2-0.0003048}-1.06429\sqrt{\al_4-2\sqrt{\beta_4}}.\nn\eea
\item The second case:
 \bea
C&=&0.5\sqrt{\al_4+2\sqrt{\beta_4}},\crn
B_2&=&0.5\sqrt{4 A^2-0.0003048}
-0.714286\sqrt{\al_4+2\sqrt{\beta_4}},\crn
m_1&=&0.5\sqrt{4 A^2-0.0003048},\,\,\,\, m_2=A, \label{case207I}\\
m_3&=&0.5\sqrt{4 A^2-0.0003048}-1.06429\sqrt{\al_4+2\sqrt{\beta_4}},\nn \eea
where
\bea
\al_4 &=&-0.00227985 + 1.76569A^2,\label{albet4}\\
\beta_4&=&1.48846\times 10^{-7}-0.00201275A^2+0.779412A^4.\nn \eea
\eit
\section{\label{K4I} The solutions with $K=0.7$ in the inverted case}
 \bit
 \item The first case:
 \bea
C&=&-0.5\sqrt{\al_4-2\sqrt{\beta_4}},\crn
B_2&=&0.5\sqrt{4 A^2-0.0003048}
-0.714286\sqrt{\al_4-2\sqrt{\beta_4}},\crn
m_1&=&0.5\sqrt{4 A^2-0.0003048},\,\,\,\, m_2=A, \label{case107pI}\\
m_3&=&0.5\sqrt{4 A^2-0.0003048}-1.06429\sqrt{\al_4-2\sqrt{\beta_4}}.\nn\eea
\item The second case:
 \bea
C&=&-0.5\sqrt{\al_4+2\sqrt{\beta_4}},\crn
B_2&=&0.5\sqrt{4 A^2-0.0003048}
-0.714286\sqrt{\al_4+2\sqrt{\beta_4}},\crn
m_1&=&0.5\sqrt{4 A^2-0.0003048},\,\,\,\, m_2=A, \label{case207pI}\\
m_3&=&0.5\sqrt{4 A^2-0.0003048}-1.06429\sqrt{\al_4+2\sqrt{\beta_4}},\nn \eea
where $\al_4, \beta_4$ are given in (\ref{albet4}).
\eit

\end{document}